\documentclass[11pt]{article}

\usepackage{amsmath, amsthm, url, epsfig}
\usepackage{amssymb, amsfonts}
\usepackage{graphicx}
\usepackage{geometry}
\usepackage{soul}
\usepackage{xcolor}
\usepackage{subfig}
\usepackage[export]{adjustbox}
\usepackage{arydshln}

\usepackage{parskip}
\usepackage{fullpage}
\title{Colorado in Context: Congressional Redistricting and Competing Fairness Criteria in Colorado}
\author{Jeanne Clelland\thanks{The first author was partially supported by a Collaboration Grant for Mathematicians from the Simons Foundation.}, 
Haley Colgate\thanks{The second author was supported by the Colorado College Summer Collaborative Research Experience.}, 
Daryl DeFord\thanks{The third author was partially supported by a Prof. Amar G. Bose Research Grant.}, Beth Malmskog, and Flavia Sancier-Barbosa}

\begin{document}

\maketitle


\begin{abstract}
In this paper, we apply techniques of ensemble analysis to understand the political baseline for Congressional representation in Colorado.  We generate a large random sample of reasonable redistricting plans and determine the partisan balance of each district using returns from state-wide elections in 2018, and analyze the 2011/2012 enacted districts in this context.  Colorado recently adopted a new framework for redistricting, creating an independent commission to draw district boundaries, prohibiting partisan bias and incumbency considerations, requiring that political boundaries (such as counties) be preserved as much as possible, and also requiring that mapmakers maximize the number of competitive districts.  We investigate the relationships between partisan outcomes, number of counties which are split, and number of competitive districts in a plan.  This paper also features two novel improvements in methodology--a more rigorous statistical framework for understanding necessary sample size, and a weighted-graph method for generating random plans which split approximately as few counties as acceptable human-drawn maps.
\end{abstract}

\section{Introduction}\label{sec:introduction}
In making an argument that a given districting plan is the result of partisan gerrymandering, we often appeal to ``the eyeball test," i.e. a sense that ugly, convoluted-looking districts are likely manipulated for someone's gain.  Similarly, non-proportional outcomes, where the percentage of seats that a party wins is very different from the percentage of the votes that party won, also strike people as particularly unfair.  However, a state's unique physical and human geography deeply affect what districts must look like and what outcomes are likely to occur even when districts are drawn without partisan bias.  For example, imagine a state in which 49\% of voters chose party A, 51\% party B.  If the party A voters are evenly distributed into every precinct in the state, then every precinct and thus every district would be majority party B, and all seats would go to party B regardless of how the district lines were drawn. The principle behind this thought experiment is played out in Massachusetts, where approximately 30-40\% of voters chose Republican candidates for U.S. House of Representatives in recent elections.  Because these Republican voters are spread fairly evenly through the state, for some elections there is probably no way of collecting enough Republican majority precincts around the state to elect even one Republican representative out of 9 that Massachusetts sends to the U.S. House \cite{MGGGMass}.  

As this unique political geography is so important, we need a way of understanding what outcomes are normal for an unbiased districting plan in a given state.  One way to do this is to generate a large ensemble of randomly created districting plans, and then use real voting data to see what election outcomes would have been under each of the randomly generated plans.  Mathematicians and computer scientists have developed theory and techniques in this area, captured under the umbrella of ``ensemble methods." The beauty of creating an ensemble is that a proposed or enacted districting plan can then be seen in context of what is possible for the state.  If the proposed plan gives an outcome that is an outlier in this context, i.e. is far from the norm, this indicates that the considered plan is very unusual for plans generated without partisan bias, providing evidence that it could be a result of partisan gerrymandering.

Statistical methods of identifying extreme gerrymanders have proven convincing in recent high-profile court cases.  For example, these methods formed part of the argument against the Pennsylvania 2011 districting map (widely considered a Republican gerrymander) in a suit brought by the League of Women Voters. The Pennsylvania Supreme Court ruled in January 2018 that the map violates the Pennsylvania constitution. These methods were also part of the March 2019 arguments to the U.S. Supreme Court against Maryland's 2011 districting map (widely considered a Democratic gerrymander). In summer 2019, the U.S. Supreme Court ruled in Rucho vs. Common Cause \cite{Rucho} that questions of partisan gerrymandering are ``beyond the reach of federal courts," refocusing the fight for fair redistricting to the levels of state and local government. Colorado voters recently passed laws that aim to create a transparent and independent process for redistricting, explicitly disallowing partisan bias in the process \cite{AmendmentY, AmendmentZ}. Colorado is thus an ideal place to explore questions of national importance and to develop best practices in redistricting. 

Our group has compiled the necessary data to apply ensemble methods to the state of Colorado.  We created a data-rich map of Colorado's voting precincts.  We used GerryChain, a tool created by the Metric Geometry and Gerrymandering Group and the Voting Rights Data Institute, to generate hundreds of thousands of random viable districting plans for Colorado.  With election data from 2018 races, we paint a picture of how likely different outcomes are for Colorado elections with these unbiased maps.  Our data and analysis provide a rich new perspective on Colorado's electoral geography.  

Colorado Amendments Y and Z, passed unanimously by both houses of the state legislature in May 2018 and by a wide margin of Colorado voters in a ballot initiative in November 2018, create independent commissions of registered voters to draw district lines for U.S. Congressional districts \cite{AmendmentY} and state legislative districts \cite{AmendmentZ}.  These laws disallow partisan bias in map drawing and require the commissions to meet many fairness criteria, including creating as many politically competitive districts as is feasible. Our group hopes to provide context to the commissioners and the staff of the state Legislative Council, who are charged with aiding the commissioners in their task.  With that goal in mind, we use ensemble analysis to present a baseline for political outcomes in Colorado's delegation to the U.S. House of Representatives.  We also consider connections between the required fairness criteria, including whether increasing the number of competitive districts in a districting plan changes the likely political outcomes for the state.

\section{Background}
\subsection{Redistricting and Federal Law}\label{subsec:federal}
A few criteria for fair maps are specified by federal law. Some points that are relevant to this analysis are outlined below.  Note that the National Conference of State Legislatures provides a brief but more comprehensive guide to Supreme Court findings in \cite{NCSLSupremeCourt}.
\begin{itemize}

\item \textbf{One Person, One Vote: }In 1962, the Supreme Court held in Baker v. Carr that U.S. House districts with unequal populations violated the Equal Protection Clause of the U.S. Constitution, creating the principle of ``one person, one vote" and creating the precedent that federal courts can have jurisdiction in cases involving state-drawn legislative plans \cite{Baker}.  The 1964 case Wesbury v. Sanders upheld the findings.  In the 1964 case Reynolds v. Sims, the court found that these principles also apply to state legislative districts.  
In 1983, the U.S. Supreme Court decided in Karcher v. Daggett that if challengers to legislative plans can show that even small deviations in population could have been reduced, then the state must prove that these variations were necessary to satisfy some legitimate goal of the state---for example, preserving political boundaries or compactness. The 2016 case Evenwel v. Abbott established that total population (as opposed to voting-age population or citizen residents) is a valid metric for complying with one person, one vote, but did not exclude other metrics.  \cite{NCSLSupremeCourt}

\item \textbf{Single Member Districts: } The 1967 federal law PL 90-196 requires that each state must establish geographic districts so that each elect a single representative to the U.S. House of Representatives.  

\item \textbf{Opportunity Districts and Racial Gerrymandering:} The Voting Rights Act (VRA) of 1965 banned vote denial or vote dilution on the basis of race, color, or language minority status \cite{VRA1965}.  Vote dilution could take the form of racial gerrymandering, which might pack minority voters into a small number of districts and/or crack minority voters into many small districts, or could occur in some at-large election schemes, which would diminish the voting power of a minority group by electing several representatives from a larger region in which the minority group's votes are not sufficient to elect any candidate.  

One result of the VRA and subsequent amendments: If it is feasible to create a districting plan so that a minority group can have the opportunity to elect a candidate of choice, and a districting plan does not do so, this districting plan can be struck down for violating Section 2 of the VRA.  \cite{NCSLSupremeCourt} 

\item \textbf{Partisan Gerrymandering is Beyond the Reach of Federal Courts:} The 2019 Supreme Court decision in the case Rucho v. Common Cause \cite{Rucho} established that ``Federal judges have no license to reallocate political power between the two major political parties, with no plausible grant of authority in the Constitution, and no legal standards to limit and direct their decisions."  The majority opinion does allow states and the U.S. Congress to pass laws restricting partisan gerrymandering, but finds that partisan gerrymandering does not violate the Equal Protection Clause of the U.S. Constitution. 
\end{itemize}

\subsection{Colorado Political History, Geography, and Election Law}\label{subsec:colorado}
Distinctive political landscape, geography, and election law make Colorado a particularly interesting place to conduct an ensemble analysis.  A western state with some larger cities and large stretches of rural area, Colorado contains large agricultural areas east of the Rocky Mountains in the Eastern Plains, ranching, ski/outdoors destinations and fruit farms on the Western Slope to the west of the continental divide, and concentrated population along the Front Range, i.e. the Interstate 25 corridor at the eastern edge of the Rockies. 
\subsubsection{Political Landscape}

Colorado is often called a ``purple" state; this perception is perhaps based on the nationally prominent statewide/national races for the Presidential electoral votes and Governor of Colorado. The numbers of U.S. Representatives from the two major parties have also shifted significantly between the two parties. To illustrate, Democratic and Republican vote shares for presidential and gubernatorial elections since 1980 are given in Table \ref{table:votes}, as well as the numbers of Representatives elected from each party. Note that the vote shares are percentages of votes for only the two major parties; votes for candidates from other parties are not counted in this table.

\begin{table}[h]
\begin{footnotesize}
\begin{center}
\begin{tabular}{l|rr|ll|cc}
Year (Race) 		& D  votes & R votes & D \% &R \% & \# D Reps  & \# R Reps  \\ \hline
1980 (Pres)               & 367973  & 652264  	&	36.07\%&	63.93\% & 3 & 2\\
1982 (Gov)                & 627960  & 392740  	&	61.52\%&	38.48\% & 3 & 3\\
1984 (Pres)               & 454974  & 821818  	&	35.63\%&	64.37\% & 2 & 4\\
1986 (Gov)                & 616325  & 434418  	&	58.66\%&	41.34\% & 3 & 3\\
1988 (Pres)               & 621453  & 728177  	&	46.05\%&	53.95\% & 3 & 3\\
1990 (Gov)                & 626032  & 358403 	&	63.59\%&	 36.41\% & 3 & 3\\
1992 (Pres)               & 629681  & 562850 	&	52.80\%&	 47.20\% & 2 & 4\\
1994 (Gov)                & 619205  & 432042 	&	58.90\%&	 41.10\% & 2 & 4\\
1996 (Pres)               & 671152  & 691848 	&	49.24\%&	 50.76\% & 2 & 4\\
1998 (Gov)                & 628846  & 633780  	&	49.80\%& 50.20\% & 2 & 4	\\
2000 (Pres)               & 738227  & 883745  	&	45.51\%&54.49\%	 & 2 & 4\\
2002 (Gov)                & 475373  & 884583  	&	34.96\%&	65.04\% & 2 & 5\\
2004 (Pres)               & 1001725 & 1101256 	&	47.63\%&	52.37\% & 3 & 4\\
2006 (Gov)                & 887986  & 625886  	&	58.66\%&	41.34\% & 4 & 3\\
2008 (Pres)               & 1288633 & 1073629 	&	54.55\%&	45.45\% & 5 & 2\\
2010 (Gov)                & 915436  & 199792  	&	82.09\%&	17.91\% & 3 & 4\\
2012 (Pres)               & 1323102 & 1185243 	&	52.75\%&	47.25\% & 3 & 4\\
2014 (Gov)                & 1006433 & 938195  	&	51.75\%&	48.25\% & 3 & 4\\
2016 (Pres)               & 1338870 & 1202484 	&	52.68\%&	47.32\% & 3 & 4\\
2018 (Gov)                & 1348888 & 1080801	&	55.52\%&	44.48\% & 4 & 3

\end{tabular}
\end{center}
\end{footnotesize}
\caption{\footnotesize Votes Cast in Colorado Elections for Two Major Parties in Top of Ticket Races and Number of U.S. Congressional Representatives from Two Major Parties, 1980-2018. }
\label{table:votes}
\end{table}

\begin{figure}[h]
\includegraphics[width=7in]{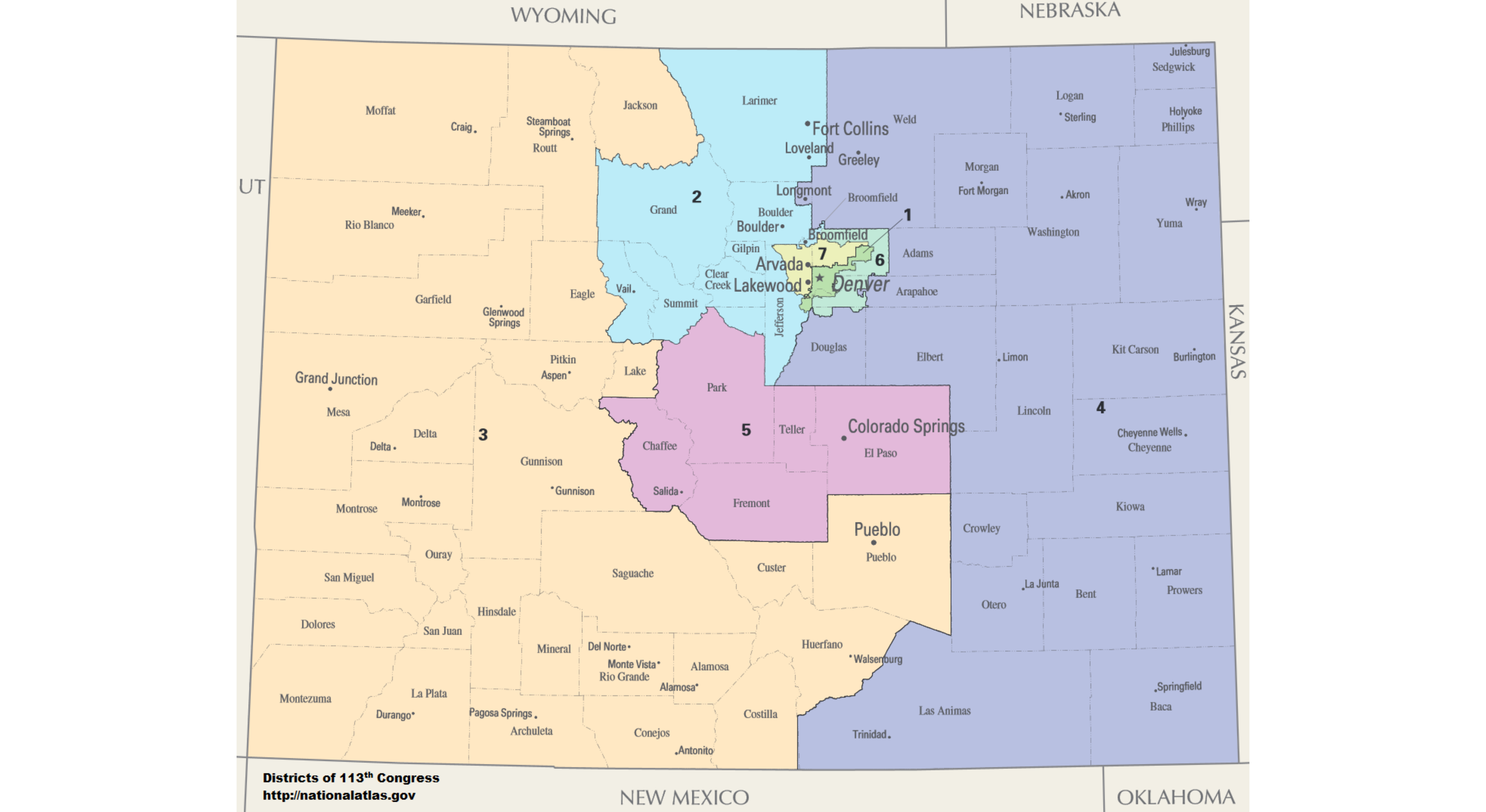}
\caption{\footnotesize Colorado's current Congressional district boundaries.}
\label{CODistrictMap}
\end{figure}
																			
Since 2002, Colorado sends 7 Representatives to the U.S. House; if population estimates are borne out by the 2020 Census, the state is projected to get 8 representatives in the next apportionment. See Figure \ref{CODistrictMap} for the current district boundaries. One of Colorado's U.S. Senators is a Democrat, one Republican. Colorado's state legislature is made up of two houses: a 35-member State Senate and a 65-member State House.  As of the 2018 election, there are 41 Democrats in the State House and 19 Democrats in the State Senate.  The 2018 election, widely considered a Democratic wave, saw all state-wide contests won by Democrats. In Table \ref{table:elections}, the vote totals for all 5 state-wide contests are given, including the raw vote totals for each contest. Note that the D and R vote shares are percentages of votes for only these two major parties, because these are the votes used in our analysis.

\begin{table}[h]
\begin{footnotesize}
\begin{center}
\begin{tabular}{l | rrr | ll}
Contest & D votes & R votes & Other votes & \% D & \% R\\ \hline
Governor & 1,348,888 &1,080,801 & 95,373 &55.52 & 44.48\\
Secretary of State & 1,313,716 & 1,113,927 & 64,992 & 54.11& 45.89\\
State Treasurer & 1,292,281 & 1,111,641 & 70,475 & 53.76 & 46.24\\
Attorney General & 1,285,464 & 1,124,757 & 81,733 & 55.64& 44.36\\
U of CO Regent (at large) & 1,246,318 & 1,031,993 & 120,714 & 54.70 & 45.30 
\end{tabular}
\end{center}
\end{footnotesize}
\caption{\footnotesize Voting Data for Statewide Offices in Colorado, 2018}
\label{table:elections}
\end{table}

As is common, rural areas in the state, including most of the Western Slope and Eastern Plains, tend to be more conservative and urban areas more liberal. However, cities in Colorado vary widely along the political spectrum; a 2014 Pew study ranked Denver, CO as the 20th most liberal big city and Colorado Springs as the 4th most conservative.  Among rural areas, the San Luis Valley in Southern Colorado leans more Democratic and contains Costilla county, one of the most Democratic-voting counties in the state.   There are also many liberal outposts among small towns on the Western Slope which are tourism/outdoor destinations, like San Miguel County, where Telluride is located.  The suburbs/exurbs of Denver are also politically divided, with some Democratic- and some Republican-leaning areas.  These dense pockets of population with regional political leanings provide fertile ground for partisan manipulation in redistricting \cite{Loevy2011}.

Colorado has a large percentage of unaffiliated voters: 40\% of active voters registered as unaffiliated as of December 2019, with 29\% registering as Democrats, 28\% as Republicans, and 3\% as other parties.  Republicans once held a plurality of registrations in the state.  In 2008, Colorado had reached a symmetry point, with about $\frac{1}{3}$ of voters registering in each of the categories unaffiliated, Democrat, and Republican.  In 2016 Colorado passed a law allowing unaffiliated voters to vote in either major party primary, which many argue has led to a rise in the percentage of unaffiliated voters \cite{frank_2019}. This would potentially make voter registration data less meaningful in determining partisan lean in Colorado than in other states.

\subsubsection{Some Points of Redistricting Law and History in Colorado}

Colorado passed a constitutional amendment in 1966 requiring single-member districts for the State Legislature (which were already required for the U.S. Congress). As of the most recent redistricting (begun in 2011), Section 47 of the Colorado Constitution specified that the districts should be contiguous, be as compact in area as possible and have the linear distance of all boundaries as short as possible, have equal population as much as possible, and strictly to deviate less than 5\% in population between most and least populous district, break county, city, and town boundaries as little as possible, and keep communities of interest in one district when possible.


The 2011 Congressional redistricting process was extremely contentious.  Control of the State Legislature was split, with the State House controlled by Republicans and the State Senate controlled by Democrats.  The two chambers were unable to agree on a set of maps for the U.S. Congressional districts by the established deadline.  Two lawsuits were filed to overturn the old, 2001 maps based on unequal population. The Denver district court and then the Colorado Supreme Court found in favor of the plaintiffs, and the court adopted a map proposed by one of the plaintiffs (the Moreno/South map).  In its decision, the court stated that ``Not a single partisan or political position or consideration played a role in this Court's task or ultimate decision"  \cite{Moreno}; however, the adopted maps were drawn by Democrats, and Republicans stated objections to the press.  It should be noted that it was not unusual for Colorado that the Legislature was unable to adopt maps in 2011 and that the courts made the final decision; the courts also made the final choice in 1981 and 2001, and significant court challenges in the 1990s forced changes in the State Legislative districts to comply with the Voting Rights Act. 


In 2018, Colorado voters and the State Legislature approved Amendments Y and Z \cite{AmendmentY, AmendmentZ} to the state Constitution, putting redistricting for both Congressional and State Legislative districts in the hands of redesigned independent commissions.  The commissions will each be made up of four Democrats, four Republicans, and four unaffiliated voters.  These members are chosen by an involved process including screening by a non-partisan panel of judges and random selection, designed to remove political agendas from the process.  The amendments also require that maps must:
\begin{itemize}
\item have contiguous districts,
\item keep district populations as equal as possible,
\item preserve communities of interest and political subdivisions,
\item minimize the number of divisions when a city, county, or town is divided, 
\item have districts as compact as reasonably possible, and
\item maximize the number of politically competitive districts.
\end{itemize}
Further, the maps must not
\begin{itemize}
\item endeavor to protect incumbent Legislators or declared candidates for the Legislature,
\item be designed to benefit any political party,
\item be designed to or result in denying any citizen the right to vote based on race or language, or dilute any minority group's electoral influence.
\end{itemize}
Amendments Y and Z do not explicitly define compactness, and competitiveness is rather strangely defined to mean ``having a reasonable potential for the party affiliation of the district's representative to change at least once between federal decennial censuses."  
%

The requirements of the laws create questions of precedence among fairness criteria.  The 2002 ``Hobbs decision" by the Colorado State Supreme Court \cite{Hobbs} found that the order of precedence for fairness criteria in legislative redistricting should be: ``(1) the Fourteenth Amendment Equal Protection Clause and the Fifteenth Amendment; (2) section 2 of the Voting Rights Act; (3) article V, section 46 (equality of population of districts in each house); (4) article V, section 47(2) (districts not to cross county lines except to meet section 46 requirements and the number of cities and towns contained in more than one district minimized); (5) article V, section 47(1) (each district to be as compact as possible and to consist of contiguous whole general election precincts); and (6) article V, section 47(3) (preservation of communities of interest within a district)."  Notice that competitiveness does not appear on this list; however, the court rulings which led to adopting the currently enacted districts refer to competitiveness as a benefit of the enacted plan in contrast to other proposed plans \cite{Moreno}. Thus competitiveness was treated as a legally valid redistricting goal before it was enshrined in Colorado law, though the definition of competitiveness has never been clear.

Going forward, however, the Legislative Council for the State of Colorado has determined \cite{Barry} that the order of precedence most consistent with the language of the new amendments is as follows: (1) population equality as required by the federal constitution; (2) compliance with the federal Voting Rights Act; (3) preserving whole communities of interest and political subdivisions; (4) compactness; (5) maximizing the number of politically competitive districts.  

%

\subsection{Ensemble Analysis}\label{subsec:ensemble}

The fundamental goal of ensemble analysis is to model the political geography of a region (in this case, the state of Colorado) in order to better understand what might be expected for a ``typical" districting plan for the state.  Plans may be evaluated with regard to a variety of measures: partisan balance of election results, geographic compactness of districts, competitiveness of district elections, preservation of communities of interest, racial/ethnic population within districts, etc.  The key idea is to create a large number of randomly generated, valid plans that satisfy all relevant legal constraints---an ``ensemble" of plans---and to use real voting data to compute the measures of interest for each plan in the ensemble.  The result is a statistical range of possible outcomes for each measure, to which any proposed plan may be compared.  If a proposed plan appears to be an extreme outlier compared to the ensemble, this may suggest that the plan was deliberately designed to achieve some specific goal, such as partisan gerrymandering.

The first issue that must be addressed is how to construct the plans in the ensemble, and how to ensure that the range of plans in the ensemble accurately represents the space of possible valid plans.  There are many different methods that have been used to construct ensembles of plans; some examples include floodfill methods (e.g., \cite{CDO00}), randomized aggregation of voting precincts into districts (e.g., \cite{CR13}), and optimization algorithms (e.g., \cite{LCW16}).
Unfortunately, the important question of how well these ensembles represent the overall space of plans is somewhat subtle and extremely difficult to answer.  The space of plans is astronomically large and poorly understood, so it is very hard to know whether a given ensemble really captures plans that accurately represent the entire space.  Furthermore, many valid plans---perhaps even most of them!---may have undesirable properties such as long, snaky districts, so it is not even clear that we would want to sample uniformly from the space of {\em all} such plans, even if we could.  It may be preferable to use a sampling method with a bias towards more ``desirable" plans, whatever that may mean. Regardless of what method is used to construct an ensemble of plans, it is important to understand how the method affects the properties of the ensemble.

For this reason, most recent work has concentrated on generating ensembles via Markov Chain Monte Carlo (MCMC) methods---basically, by taking a random walk on the space of valid plans.  (See, e.g., \cite{HKLVBRM18}, \cite{MV14}.)  In these methods, an ensemble is generated by starting from a particular plan (a ``seed plan") and modifying it in some way to create a new plan.  If the new plan is valid, it is added to the ensemble and the process is repeated, with the new plan as the starting point.  MCMC sampling methods have a well-developed theory and a long history of applications (see, e.g., \cite{Diaconis09}).  Most importantly, a sufficiently long Markov chain is theoretically guaranteed to produce a sample that accurately represents a specific probability distribution on the entire space.  The main issues that need to be addressed are: 
\begin{itemize}
\item How well can we characterize the probability distribution that we are sampling from?  This distribution is determined by the method used to move from one state to another, but not in a way that is easily understood.
\item How long does the chain need to be in order to guarantee a representative sample?  This is known as the ``mixing time" for the chain; currently there is no known way to determine it rigorously, and heuristic methods must be used to determine experimentally when a chain appears to be long enough.
\end{itemize}

\begin{figure}[h]
\begin{center}
\includegraphics[width=6in]{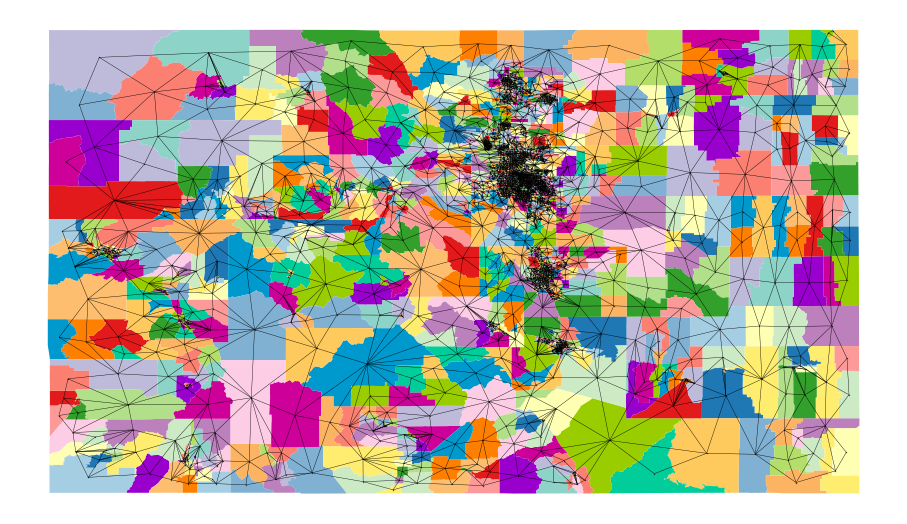}
\end{center}
\caption{Precinct map and dual graph for Colorado}
\label{CO-dual-graph}
\end{figure}

For this type of analysis, it is natural to build districts from voting precincts, as these are the smallest geographic units for which voting data is readily available.  We begin by constructing the ``dual graph" for the state, consisting of a vertex for each voting precinct, with two vertices connected by an edge whenever the corresponding precincts share a geographical boundary.  (See Figure \ref{CO-dual-graph}.)
Data for each precinct (e.g., population, election results) is associated with the corresponding vertex, and a districting plan is represented by a partition of the graph into connected subgraphs, each representing a district.

The most important algorithmic choice is how to move from one districting plan to the next in the Markov chain.  The algorithm used to generate a new plan from a given one is referred to as the ``proposal function," and it may or may not be accompanied by an ``acceptance function" that determines whether (with some probability) a proposed plan is accepted as the next step in the Markov chain.

The first applications of MCMC methods to the sampling of district plans mostly used versions of a proposal function that we refer to as a ``Flip step."  In its simplest form, a randomly selected precinct on the boundary between two districts is flipped from its current district to its neighboring district to create a new plan.  If the resulting plan passes validity checks (e.g., if all district populations remain equal to within some predetermined error tolerance), then the new plan is accepted (possibly with some predetermined probability) as the next step in the Markov chain.  Variations of this method have been used by Mattingly, et. al. to analyze Congressional districting plans for North Carolina (e.g., in \cite{HKLVBRM18}); this work has been cited in a variety of important litigation, including the 2019 U.S. Supreme Court case Rucho v. Common Cause \cite{Rucho}.  

Advantages of this method include its simplicity---each step in the chain is computationally cheap and very fast to compute---and it can be implemented in such a way that the sampling distribution is the uniform distribution on the space of plans. (See, e.g., \cite{Chikina2860}, \cite{DDS19}.)
But it has significant disadvantages as well: Without additional embellishments, it tends to create very snaky, non-compact districts that are atypical of human-drawn plans.  More importantly, experimental evidence suggests that mixing times for Flip chains are extremely long; even the longest chains that have been used in practice appear to retain recognizable characteristics of the starting plan, so it is difficult to argue that a given chain accurately represents the entire sample space.  This observation is supported by the work of Najt, DeFord, and Solomon in \cite{najt2019complexity}, where it is shown that sampling from the uniform distribution via {\em any} method is generically hard.  Nevertheless, this method does have useful applications; most notably, as in the work of Mattingly, et. al. in \cite{HKLVBRM18}, ensembles generated by this method can be useful for exploring the space of districting plans that are ``close to" a particular plan.  If a proposed plan is an extreme outlier for a Flip ensemble, it may suggest that the district boundaries were precisely engineered to achieve a specific result.

In \cite{VA-report}, the Metric Geometry and Gerrymandering Group introduced an alternate proposal function that they called ``ReCom," inspired by the idea of recombinant DNA; this method was developed further in \cite{DDS19}.  In this method, a pair of adjoining districts is selected at random and merged to create a double-sized district.  This double district is then re-partitioned into a pair of districts using a spanning tree method: A variation of Kruskal's algorithm is used to create a random spanning tree for its dual graph, and then an edge is selected to cut the tree into two districts with approximately equal population.  (If no such edge exists, then the tree is discarded and a new spanning tree is drawn.)
With this method, each step in the Markov chain is much more computationally intensive than each step in a Flip chain, but this extra cost is more than made up for by making much larger changes to the districts at each step.  In \cite{VA-report}, a variety of statistics for chains starting from random seeds exhibited vastly better mixing properties with a 20,000 step ReCom chain than with a 10 million step Flip chain.  

It was also observed in \cite{VA-report} that districts generated by a ReCom chain tend to be much more compact than districts generated by a Flip chain.  This observation is related to the fact that the probability distribution associated to a ReCom chain is closely related to the number of spanning trees possessed by the dual graphs of the districts in the plans, and more compact districts generally have many more spanning trees than less compact districts.  While this distribution is definitely not uniform on the entire space of districting plans, its bias towards more compact districts is arguably a desirable feature: Enacted plans are generally expected to have relatively compact districts, and if the goal is to create a sample of plans similar to those that might be drawn by a human, then a sampling algorithm that prefers more compact plans may be seen as a good starting point.

\subsection{Synthesis and Focus Questions}\label{subsec:synthesis}
We seek to use ensemble analysis to shed light on Colorado's political landscape, put the current districting plan for the state in context, and consider the consequences of and relationships between fairness criteria in Colorado law.  This requires a set of choices about how to interpret these criteria and how to set parameters for our Markov chain to generate a reasonable ensemble of plans.  

\begin{itemize}

\item \textbf{Contiguity} We consider districts made up of whole voting precincts. A district is considered contiguous if the induced subgraph of the precincts comprising the district within the dual graph of precincts for the state is connected.  

\item \textbf{Equal Population} We use the 2010 Census population for our analyses, because we seek to put the enacted plan in context and this was the data used to generate those districts.  The current districting plan includes a population deviation of 245 people between the least and most populous district.  In our Markov chain, we allow a district population deviation of at most 1\% from the ideal district size. 
This error tolerance was chosen because the MCMC algorithm requires that some deviation be allowed, and the 1\% threshold has become common practice for modeling Congressional districts when using precincts as building blocks.  In practice, mapmakers can and do split precincts in order to balance district populations more accurately.  

\item \textbf{Compactness}  Amendments Y and Z do not specify any one measure of compactness for future redistricting.  We follow Colorado precedent of considering the total perimeter of a districting plan.  We therefore will seek to generate an ensemble of plans with total perimeter reasonably close to that of the 2012 enacted plan. 
This consideration contributed to our choice to use a ReCom instead of a Flip, as ReCom tends to produce relatively compact districts without further tinkering with the Markov chain.  

\item \textbf{Preserving Political Boundaries}  There are an enormous number of potential political subdivisions that could be considered in drawing a map, and it is not feasible to preserve all political boundaries.  In this analysis we focus on county boundaries, as most cities and many other political entities in the state are contained within one county.  We consider the number of counties which are divided between two or more districts, as well as the total number of splits (which, unlike the former measure, increases if one county is divided many times).  This paper introduces a new weighted graph method of reducing the number of county splits, creating an ensemble with a range of county splits comparing reasonably to the number of splits in the enacted plan.

\item \textbf{Partisan Bias} We seek to identify partisan bias by creating an ensemble of potential maps, then using voting data from the 2018 general election to compute several partisan measures.  First, we determine the partisan balance of each district using the voting data from several statewide elections.  In this analysis, we included State Treasurer, Secretary of State, and gubernatorial races.  Though we are modeling U.S. House elections, we do not use U.S. House voting data because incumbency effects will vary across the state.  We chose the gubernatorial race because it had the largest number of votes among statewide races in 2018 (2,535,062 votes), with the Secretary of State race in second (2,492,635 votes).  We included the State Treasurer's race because it also had a large number of votes (2,474,397), and the voting data from the 2010 State Treasurer's race had been used by redistricting commission Chair Mario Carerra in his analyses of competitiveness in the 2011 Colorado state legislature redistricting process \cite{Loevy2011}.  Based on the partisan balance of each district with the voting data from the three contests, we are able to look for evidence of partisan bias in the relationship of the enacted plan to the ensemble.  

\item \textbf{Competitiveness} We adopt the ``vote-band" method of measuring competitiveness \cite{DDS20}, in which a district is considered competitive if Democrats and Republicans both receive 45\% to 55\% of the combined Democratic and Republican votes (votes for candidates of other parties are not considered).  This was the definition of competitiveness adopted by Carerra in 2011 \cite{Loevy2011}, though other definitions based on voter registration (not historical voting data) were considered by Commissioner Loevy and by the Colorado State Supreme Court in their 2012 ruling on the Colorado Congressional redistricting \cite{Loevy2011, Moreno}. Definitions of competitiveness relying on voter registration would be less useful in Colorado since 2016, when unaffiliated voters have been able to vote in either major party primary, removing a major incentive for party registration. Other definitions of competitiveness abound and could be considered under Colorado law, which is extremely broad.\footnote{We note that \cite{DDS20} also examines Colorado's current legal definition of competitiveness, that is, ``having a reasonable potential for the party affiliation of the district's representative to change at least once between federal decennial censuses"\cite{AmendmentY} using probability, and finds that a literal reading of this law might be that ``any district in which both parties have at least a 13\% projected chance of winning might match the Colorado law, since $(1-0.13)^5 \approx 0.5$" \cite{DDS20}.} We believe that the simple vote-band model requires the fewest assumptions and will be most satisfactory for this analysis. However, we do consider that we may miscount competitive districts by this model, based on the fact that the state-wide vote shares are at the edge of the competitive band. 
Thus we also determine how many districts would fall into the competitive vote band in each plan if the state-wide vote shares had been 50\% for Democrats and Republicans, under an assumption of uniform partisan swing.

\item \textbf{Communities of Interest} Communities of interest are complicated and the questions of definition, geographic boundary, and precedence are too nuanced to be captured in our current model.  Thus we do not claim to create an ensemble of maps which preserve communities of interest in any meaningful way. Some have argued (see \cite{Loevy2011} for a Colorado example) that claims to preserve communities of interest are veils for partisan manipulation. The fact that our analysis does not include communities of interest allows us to explore the validity of these criticisms---does the enacted plan, which was approved by the Denver District Court based upon extensive consideration of communities of interest, give Democrats more seats than an ensemble of plans drawn without considering communities of interest?

\item \textbf{Voting Rights Act}  In the 2011 ruling that approved the enacted Congressional maps, the Denver District Court found that at the time there was not sufficient density of any minority group to draw an opportunity district in Colorado \cite{Moreno}, so we do not include constraints related to opportunity districts in our Markov Chain. 

\end{itemize}

Focus questions for our study:

\begin{enumerate}

\item \textbf{What is the baseline?} We estimate a range of outcomes that Coloradans could expect from a redistricting plan drawn without partisan bias.

\item \textbf{Is the current plan fair?} We consider the enacted plan in context of the ensemble, to determine whether there is evidence that the enacted plan benefits Democrats, as some Republicans alleged at the time it was adopted.  

\item \textbf{How does the number of county splits relate to partisan election outcomes?} 

\item \textbf{How does the number of competitive districts relate to partisan election outcomes?}

\item \textbf{How does the number of competitive districts relate to the number of county splits?}

\end{enumerate}

\section{Methods}\label{sec:methods}
In this section, we discuss the implementation details of our ensemble generation process.  This includes the construction of the underlying dual graph in \ref{ssec:CO_Dual}, the Markov chain sampling procedure \ref{ssec:recom}, the statistical analysis that informed the choice of weight value in \ref{ssec:weighted}, and the statistical framework used to determine sample size in \ref{ssec:SampleSize}. 

Computational experiments were run on a combination of Google's CoCalc platform and an Ubuntu 16.04 machine with 64 GB memory and an Intel Xeon Gold 6136 CPU (3.00GHz) processor. Ensembles were generated with the 0.2.12 release of the open-source GerryChain software.
The main experimental results reported in Section \ref{sec:data} are based on an ensemble of size 2,000,000, generated with the weight parameter of 20 as discussed below. 

\subsection{Data Processing} 
\label{ssec:CO_Dual}
The underlying geographic units used to form the dual graph for our analysis are the precincts used in the 2018 general election, shown in Figure \ref{CO-dual-graph}.  Our precinct map relied on a spring 2018 map compiled by Todd Bleess at the Colorado State Demography Office, who had collected the precinct maps to share with the US Census Bureau. Our team updated the maps to match the November 2018 boundaries by comparing the Bleess maps to voter files from November 2018.  If there were new precincts listed, the new maps were obtained from the county when possible.  In some cases new maps were not available, so boundaries were inferred by geolocating the addresses of registered voters in the publicly available state voter file. The projection type for our GIS shape files was  NAD83.

In creating the precinct graph, we chose to merge certain precincts which we determined no reasonable mapmaker would split.  We immediately merged precincts Elbert 1 and Elbert 3, because Elbert 1 is the city of Elizabethtown, which fully surrounds parcels of non-incorporated land, technically within Elbert 3. Thus any plan with contiguous districts would require these two precincts to remain linked. We then determined that 14 counties had a population below 5000 in 2010, which is less than $0.1\%$ of the state population, making it unlikely that these counties would ever need to be split for population equality.  Therefore we merged the precincts within those counties. Three small counties already have only one precinct: San Juan, Mineral, and Hinsdale. The counties with multiple precincts that we merged are Jackson, Kiowa, Cheyenne, Dolores, Sedgwick, Costilla, Baca, Custer, Ouray, Phillips, and Washington.

Precinct-level election data from the 2018 general election was obtained from the Colorado Secretary of State's office, and associated with the appropriate vertices of the dual graph \cite{election_data}.  In our analysis, we consider only votes for Democratic and Republican candidates; votes for other candidates are not counted in any total. Population and demographic data from the 2010 Census was downloaded at the census block level \cite{2010_census}. Since census blocks sometimes  cross precinct boundaries, we used GIS technology to determine the centroid of each census block, and associated the entire population of the block to the vertex of the precinct containing the centroid.  We also used GIS to calculate the length of each boundary between precincts and associated this data to the corresponding edges. 

\subsection{Markov Chains on Districting Plans}
\label{ssec:recom}

The Markov chains used to generate the ensembles in this paper are a variation of the ReCom method described in \cite{DDS19}. Starting from an initial plan, at every step the subgraph consisting of the nodes belonging to two adjacent districts are merged and a spanning tree is constructed on this subgraph. The spanning tree is generated by assigning a random weight to each edge and then using Kruskal's algorithm to select a spanning tree of maximal weight. New random weights are assigned to each edge at every iteration; we discuss the choice of random function for selecting the edge weights in Section \ref{ssec:weighted} below. Once a spanning tree is constructed, an edge from the tree is removed and the two remaining connected components become the proposed districts for the next step. The edge that is removed is selected so that all of the districts are within 1\% of the ideal population. 

\begin{figure}\begin{center}
\includegraphics[width=5in]{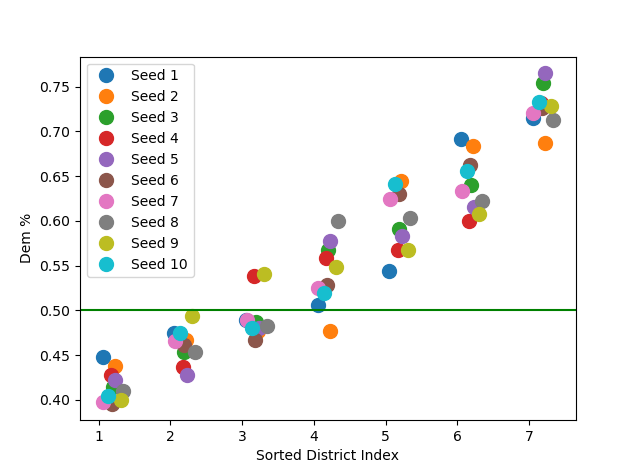}

\includegraphics[width=5in]{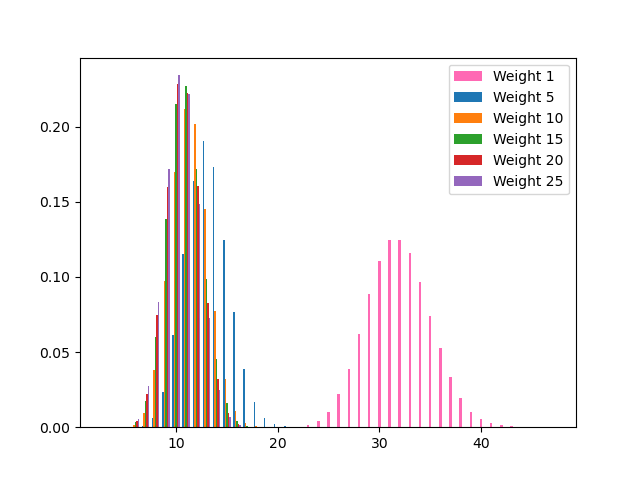}
\end{center}

\caption{\footnotesize The upper plot shows a comparison of Democratic vote share vectors for the different seeds used to start the ensembles.  The lower plot shows a comparison of the number of split counties as a function of the weight parameter in Kruskal's algorithm. When the parameter is set to 1, the algorithm splits many more counties (right, hot pink) than the versions with parameters between 5 and 25. The main ensembles analyzed in this paper use a weight of 20, chosen to balance computational performance with county preservation.}
\label{fig:seeds}
\end{figure}

The initial seeds for the comparison runs were generated with the recursive spanning tree method described in \cite{DDS19}. Figure \ref{fig:seeds} shows examples of these seeds and their partisan statistics, demonstrating that the chains began at a diverse collection of starting points. 

\subsection{County Weighted Trees}
\label{ssec:weighted}

In the default ReCom algorithm implemented in GerryChain, the edge weights assigned at each step are drawn independently and identically distributed (i.i.d.) from the uniform distribution on $[0,1]$. While the plans generated by this approach are contiguous, population balanced, and frequently have compactness scores comparable to those of human-drawn plans, they do not incorporate any information about county boundaries and tend to split many more counties than would likely be permitted in an enacted plan. 

\begin{figure}
\includegraphics[width=.45\textwidth]{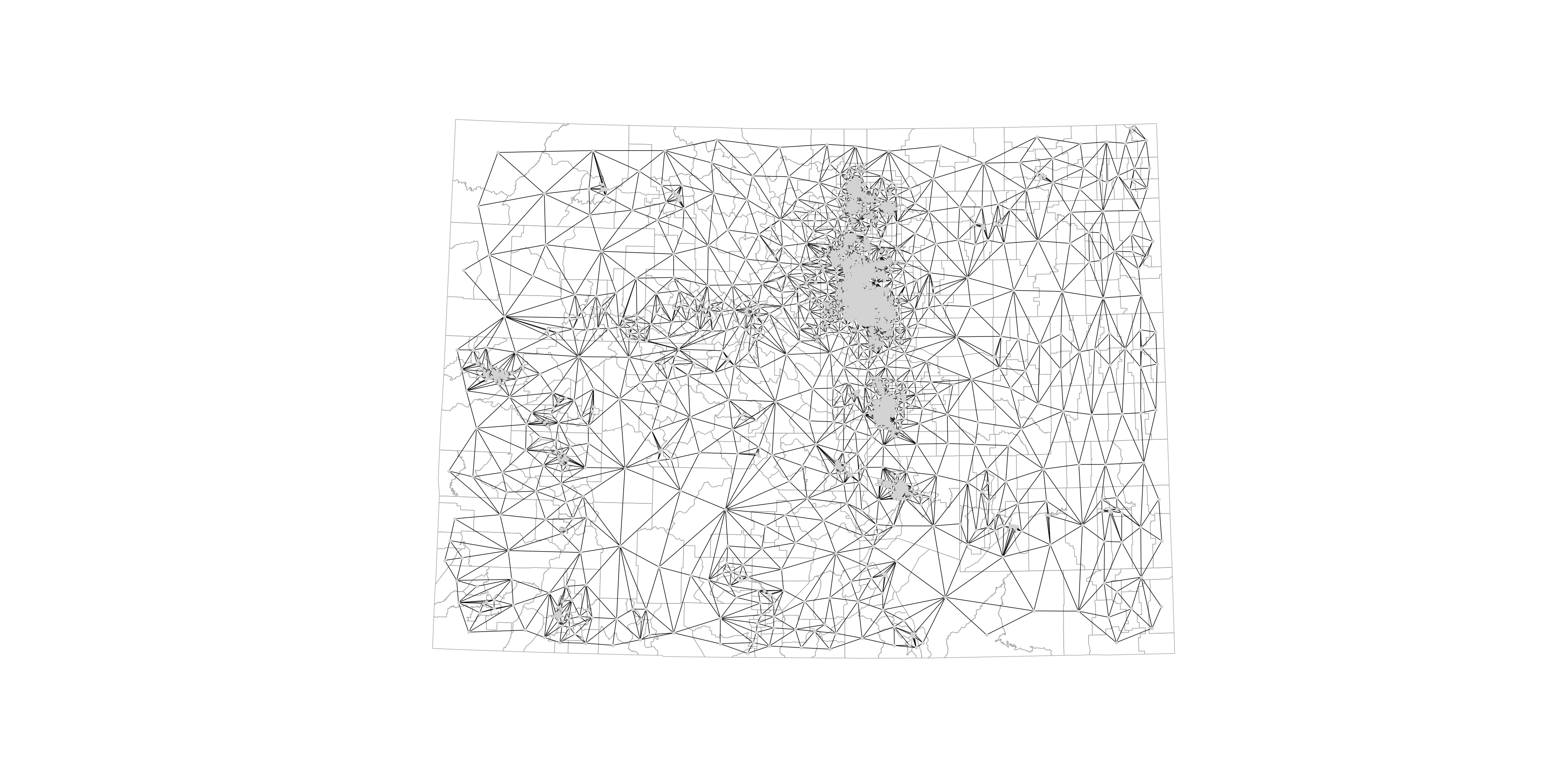}
\includegraphics[width=.45\textwidth]{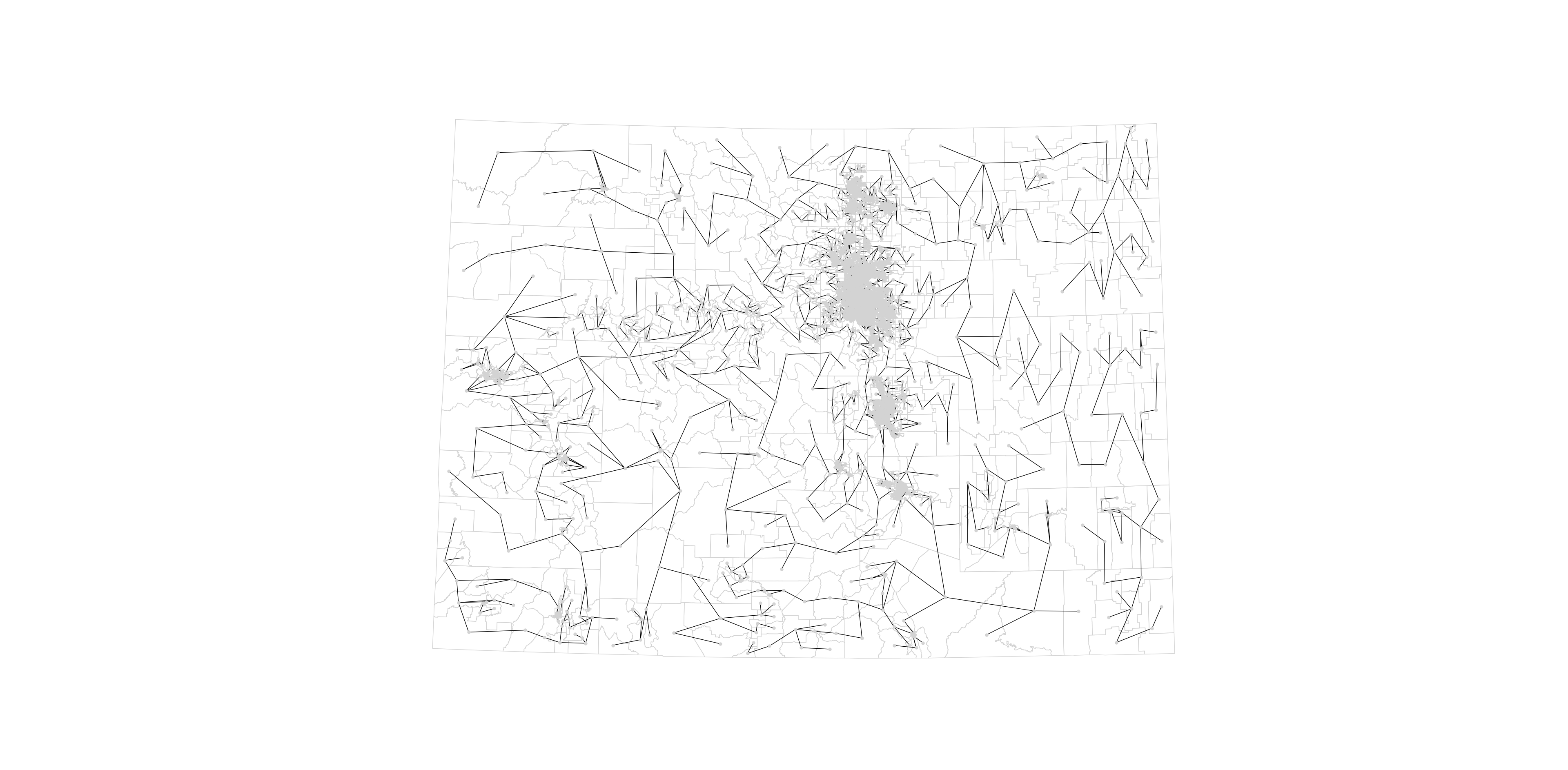}\\
\includegraphics[width=.45\textwidth]{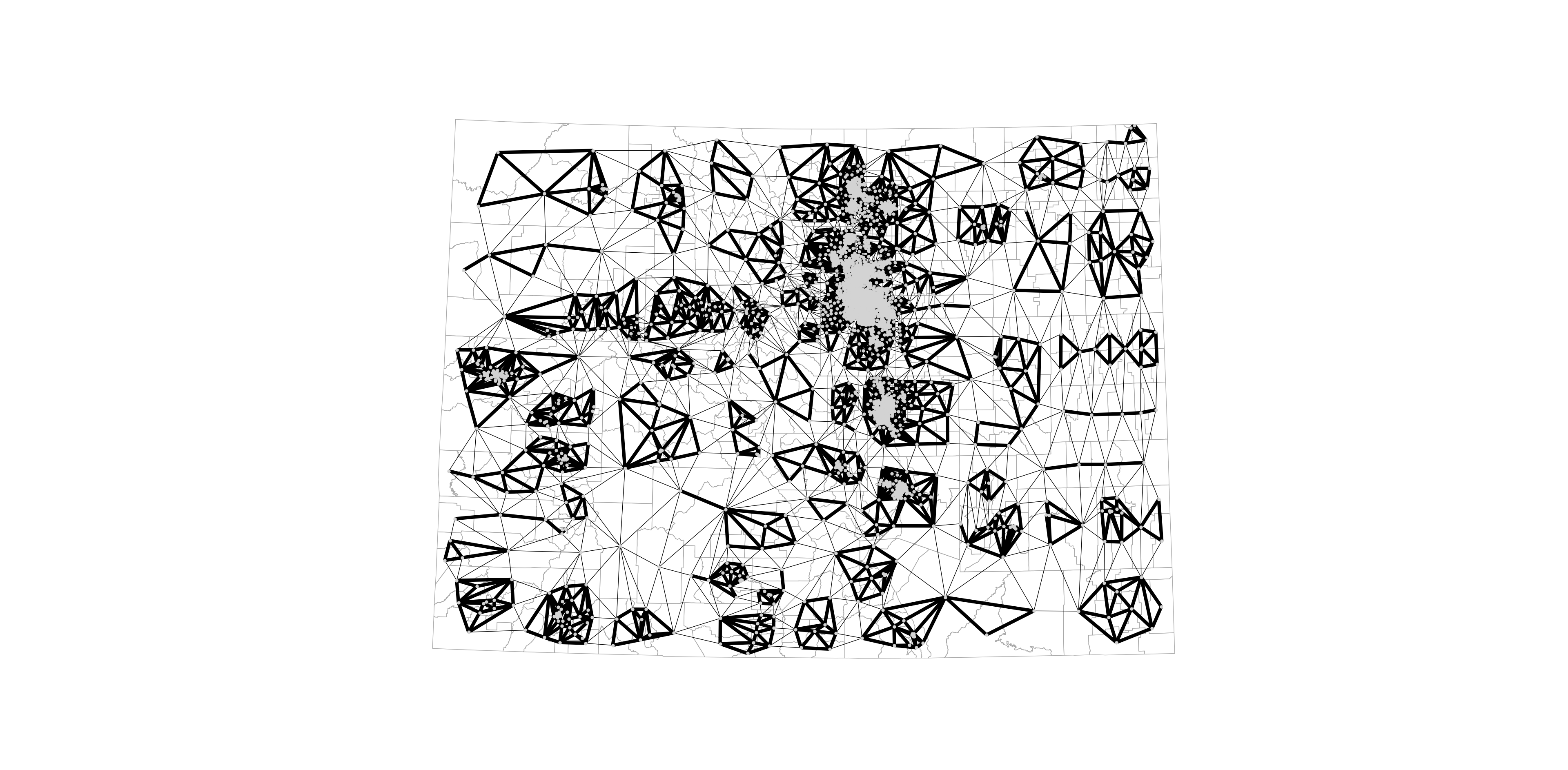}
\includegraphics[width=.45\textwidth]{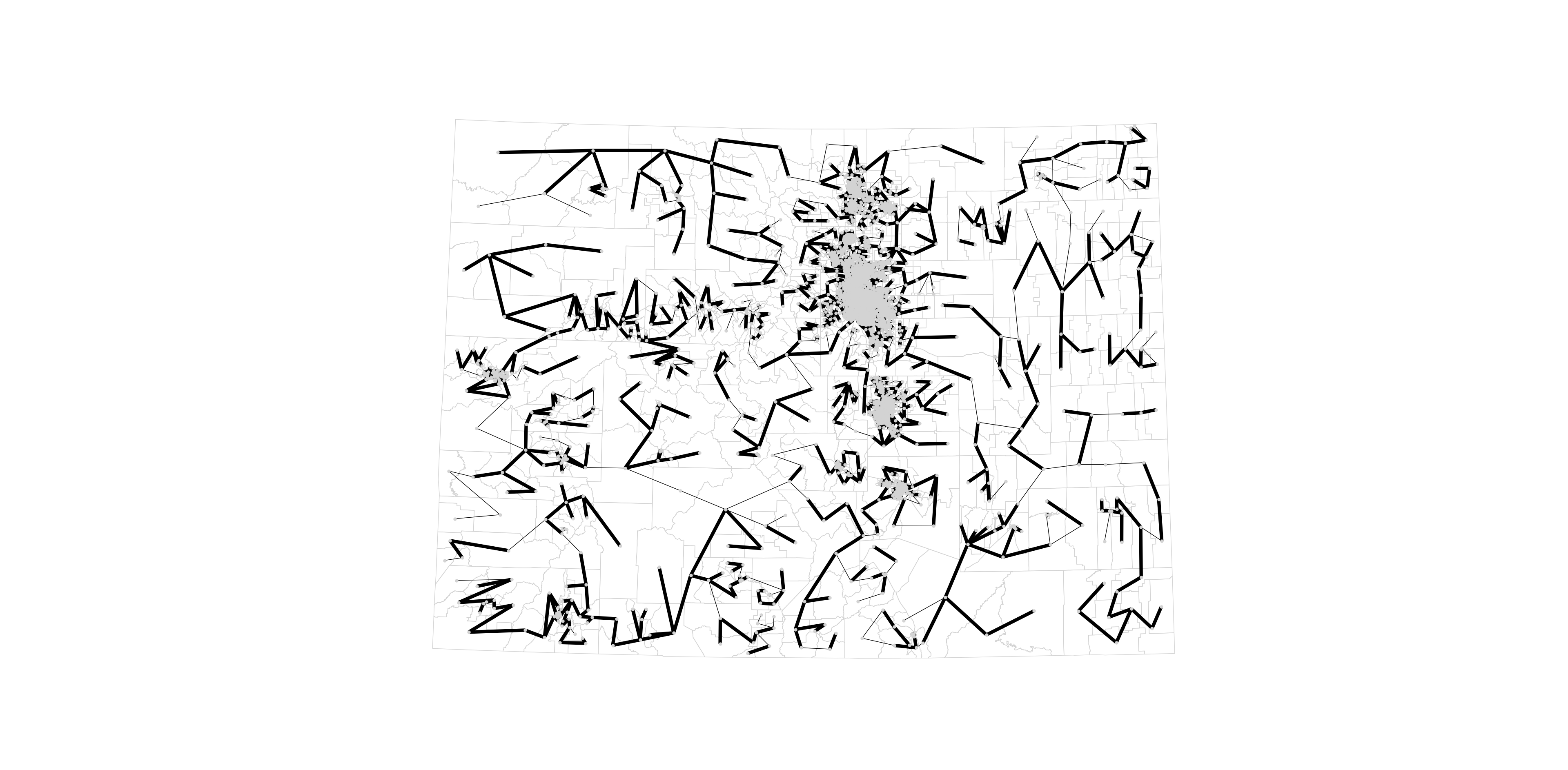}

\caption{\footnotesize Schematic demonstrating the county-weighting procedure. The top row shows the dual graph edges with equal weights and an associated spanning tree, while the bottom row shows upweighted edges between nodes belonging to the same county. Highly weighted edges are more likely to be added to the final spanning tree; the bottom right plot demonstrates an example of this behavior.  }
\label{fig:weight_schematic}
\end{figure}

In order to sample from plans that split fewer counties, we replaced the weighting function in the proposal method with one that increased the weights on edges between nodes that belong to the same county by multiplying the weight values by a constant factor. That is, the weights for intra-county edges are drawn i.i.d. 
 uniformly on $[0,w]$, where $w$ is a weight parameter, while the weights for inter-county edges are drawn from $[0,1]$ as in the standard version. See Figure \ref{fig:weight_schematic} for a schematic demonstrating this adjustment. 
  This upweighting means that more intra-county edges are included in the final spanning tree, while the spanning trees are still randomly constructed.  Indeed, since the two weight distributions overlap, this does not change the support of the state space for the associated walk, although it shifts the distribution significantly towards plans with fewer county splits. 

In order to select a weight parameter $w$  for our experiments, we compared the performance of several possibilities. Figure \ref{fig:seeds} summarizes these results, showing histograms of the number of observed splits in chains of length 100,000 using $w\in\{1,5,10,15,20,25\}$. There did not appear to be  significant differences for weights greater than 10 on the observed numbers of splits, so we somewhat arbitrarily selected $w=20$ as a reasonable choice for the main ensemble analyzed in Section \ref{sec:data}. Fine tuning of this parameter is an area for future work.



\subsection{Monte Carlo sample size}
\label{ssec:SampleSize}

Since neither mixing time nor ergodicity of the chain can be determined with certainty, we must rely on heuristic convergence diagnostics. Each measure of interest (for example, district vote share, seat share, 
etc.) may have its own rate of convergence to a stationary distribution, and therefore the number of runs must be long enough to achieve satisfactory convergence diagnostics for all measures of interest. 
In each created map, the seven generated districts can be ranked from least to greatest Democratic vote share, and we considered convergence diagnostics for each ranking category (least Democratic, second-ranked, etc.). We also looked at convergence diagnostics for the number of Democratic seats to check how these differed from those for district vote share. To set up the proportion of Democratic votes at the precinct level, we used data from three statewide races in 2018: Governor, Treasurer, and Secretary of state. We run diagnostics using each of these races.

The overall plan for obtaining a reasonable Monte Carlo sample size $n$ was guided by the following heuristics:

\begin{enumerate}
	\item The length of the chain should be a large multiple of the time it takes the autocorrelations of the measures of interest to decay to nearly zero. This ensures that the chain cycles through the distribution enough times to give us the equivalent of many independent draws.
	\item The length of the chain should be large enough so that the empirical distributions of the measures of interest obtained from one run of the chain are nearly indistinguishable (according to some measure) from the distributions obtained from another run. 
\end{enumerate}

The first item uses within-chain information, while the second uses between-chains information to ensure that our ensembles produce distributions that are adequately independent of starting point. Naturally, these guidelines cannot detect pseudo-convergence, but in the case of convergence, they increase confidence that we have run the chain long enough to achieve mixing.

\begin{figure}[h!]
	\centering
	\subfloat[]{
		\includegraphics[width=8cm]{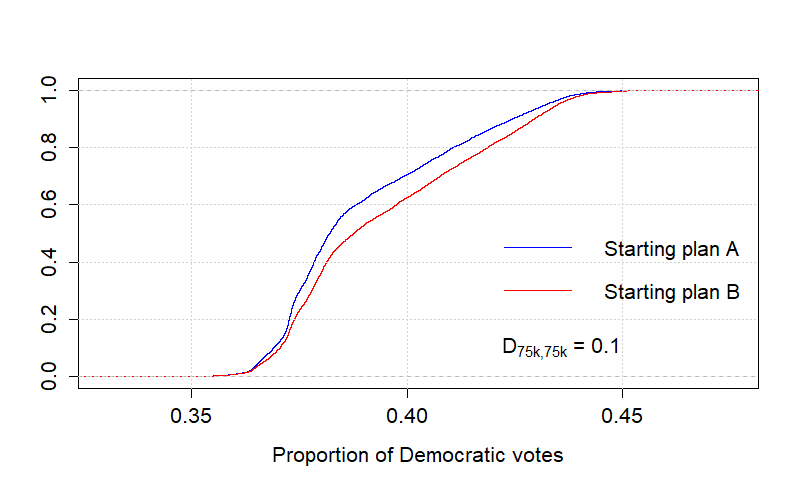}
	}
	\subfloat[]{
		\includegraphics[width=8cm]{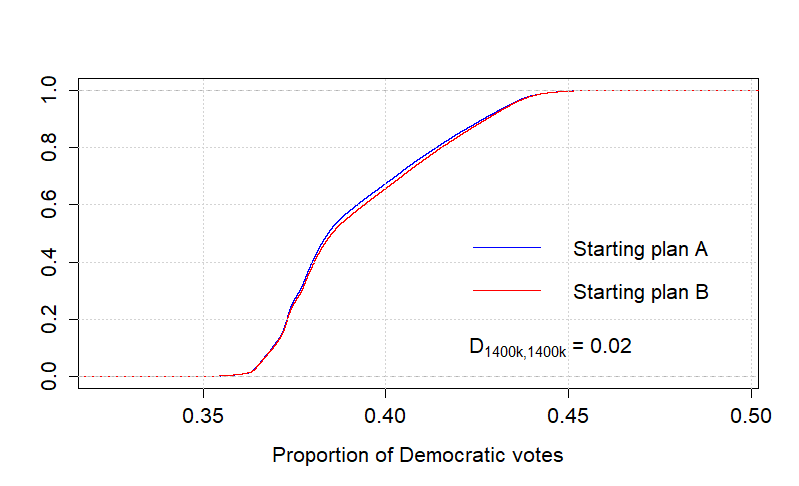}
	}
	\hspace{0mm}
	\subfloat[]{
		\includegraphics[width=8cm]{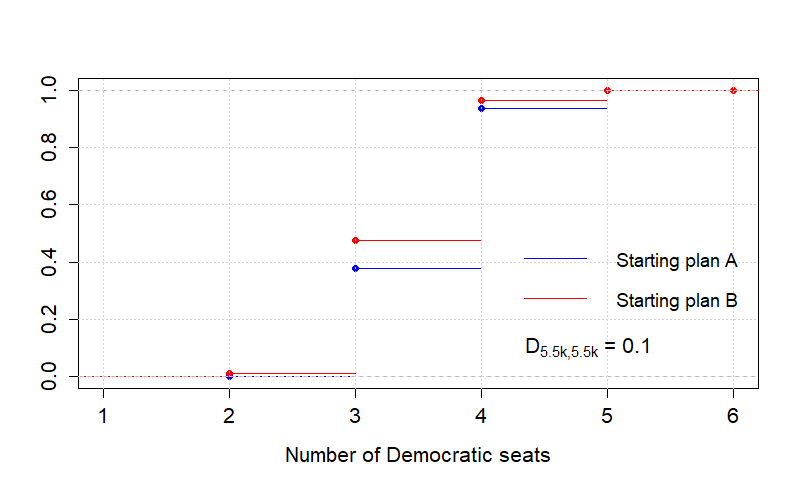}
	}
	\subfloat[]{
		\includegraphics[width=8cm]{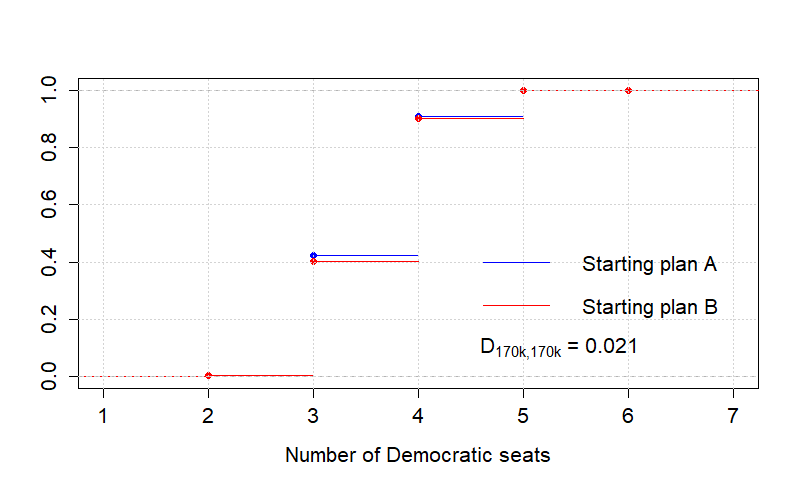}
	}
	\caption{\footnotesize ECDFs for Monte Carlo samples of various sizes (starting with plans A and B) for the vote share of the least Democratic district and the number of Democratic seats, using precinct data from the 2018 Secretary of State's race and a weight factor of 20. (a) Two ECDFs for the vote share of the least Democratic district; $D_{75k,75k} \approx 0.1.$ (b) Two ECDFs for the vote share of the least Democratic district; $D_{1400k,1400k} \approx 0.02.$ (c) Two ECDFs for the number of Democratic seats; $D_{5.5k,5.5k} \approx 0.1.$ (d) Two ECDFs for the number of Democratic seats; $D_{170k,170k} \approx 0.02.$ }
	\label{fig:twoECDFs}
\end{figure}

The measure used to test how close two empirical distributions are from one another was the two-sample Kolmogorov-Smirnov (KS) statistic. The two-sample KS statistic, $D_{m,n}$, is the maximum vertical distance between two empirical cumulative distribution functions (ECDFs) derived from two samples of sizes $m$ and $n$. Figure \ref{fig:twoECDFs} shows the ECDFs for Monte Carlo samples of various sizes (starting on distinct plans, here called A and B) for the vote share of the least Democratic district and for the number of Democratic seats, using precinct data from the 2018 Secretary of State's race and a weight factor of 20. 

\textbf{Sample size criteria}: We will consider Monte Carlo samples of size $n$ to be large enough if $n$ is such that $E(D_{n,n}) \leq 0.01$ and $n$ is a large multiple (by a factor of at least 1000) of the time it takes the autocorrelations of the measures of interest to decay to a value smaller than or equal to 0.01. In the case of district vote shares, we look at the time it takes the slowest decaying autocorrelation, out of the seven ranked congressional vote shares, to decay to 0.01 or less.

Smirnov's theorem \cite{smirnov_1939} states that for two independent and identically distributed (i.i.d.) samples drawn from the same continuous distribution, as $n$ grows, $\sqrt{\frac{n}{2}}D_{n,n}$ approaches the Kolmogorov distribution $K = \sup_t B(t)$, where $B$ is a Brownian Bridge. That is, $K$ does not depend on the distribution that the samples are drawn from. In that case, $E(D_{n,n}) \approx \sqrt{\frac{\pi}{n}}\ln 2 \approx \frac{1.23}{\sqrt{n}}$, and the quantiles of $D_{n,n}$ are also asymptotically proportional to $\frac{1}{\sqrt{n}}.$ (See, e.g., \cite{DKT20}.) In particular,  for  i.i.d. samples (from continuous distributions) of size at least 15,094, it follows that $E(D_{n,n}) \leq 0.01$. However, our MCMC samples are not independent and are only identically distributed after a stationary distribution has been reached, which implies that they will need to be larger than 15,094 (likely much larger) to ensure that $E(D_{n,n}) \leq 0.01$. When i.i.d. samples are discrete, the limiting distribution of $\sqrt{\frac{n}{2}}D_{n,n}$ is no longer independent of the distribution that the samples are drawn from; nevertheless, $E(D_{n,n})$ and the quantiles of $D_{n,n}$ are still asymptotically proportional to $\frac{1}{\sqrt{n}}.$ Therefore, we anticipated that for the measures of interest in our ensembles, the expectation and quantiles of $D_{n,n}$ would also be asymptotically proportional to a power function of $n$. 
We used empirical approximations of $E(D_{n,n})$ for various values of $n$ to estimate appropriate sample sizes, and empirical approximations of the 0.05 and 0.95 quantiles of $D_{n,n},$ denoted by $Q_{0.05}(D_{n,n})$ and $Q_{0.95}(D_{n,n}),$ to construct prediction intervals for $D_{n,n}.$  

To get estimates for $E(D_{n,n})$, $Q_{0.05}(D_{n,n})$ and $Q_{0.95}(D_{n,n})$
for our non-i.i.d. ensembles, we used independent runs of the chain starting from ten different plans (labeled A through J). To compare rates of convergence for ``weighted" versus ``unweighted" chains, we produced estimates for chains with weight 1 and 20. We ran the chains for 2 million steps and recorded $D_{n,n}$ for each pair of ensembles for several values of $n$. More specifically, we recorded ${10 \choose 2} = 45$ values of $D_{n,n}$ for each $n$. We then estimated $E(D_{n,n})$, $Q_{0.05}(D_{n,n}),$ and $Q_{0.95}(D_{n,n})$ for 47 values of $n$ ranging from 100,000 to 2 million. 
A few estimates were done for values of $n$ below 100,000 for graphing purposes only; they were not used for fitting the power functions that related $n$ with $E(D_{n,n})$, $Q_{0.05}(D_{n,n})$ and $Q_{0.95}(D_{n,n})$. In particular, for a given value of $n$, the estimate for $E(D_{n,n})$ was the average of all 45 calculated realizations of $D_{n,n}$, while the estimates for $Q_{0.05}(D_{n,n})$ and $Q_{0.95}(D_{n,n})$ were the sample quantiles of the 45 realizations of $D_{n,n}$. This was done for each of the seven ranked Democratic voting shares and number of seats, for three races in 2018. 

\begin{figure}[h!]
	\begin{center}
		\includegraphics[width=4.5in]{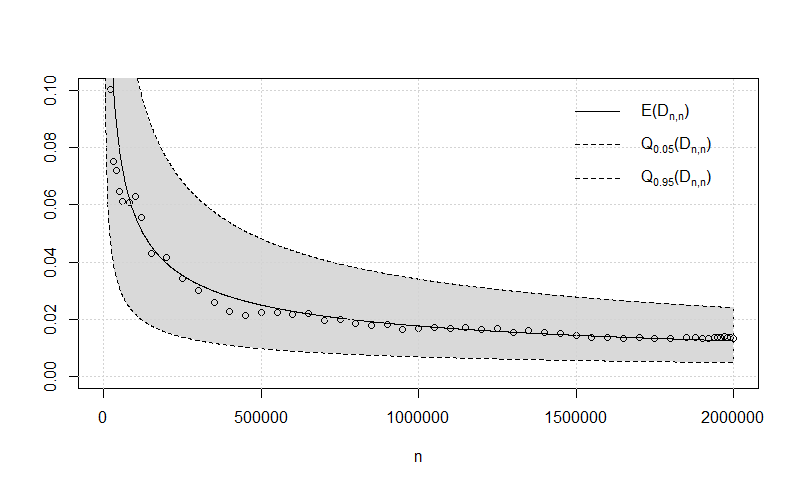}
	\end{center}
	\caption{\footnotesize Approximate relationship between $E(D_{n,n})$ and $n$ for the vote shares of the least Democratic district, using precinct data from the 2018 Treasurer's race and a weight factor of 20.  The graph also shows a 90\% prediction interval for $D_{n,n}$, which uses the estimated 0.05 and 0.95 quantiles for $D_{n,n}$. In this case, $E(D_{n,n}) \approx \frac{17.65}{\sqrt{n}},$ $Q_{0.05}(D_{n,n}) \approx \frac{6.88}{\sqrt{n}},$ and $Q_{0.95}(D_{n,n}) \approx \frac{34.02}{\sqrt{n}}.$}
	\label{fig:fitn}
\end{figure}

After the estimates described above were made, we fitted a power function that related $E(D_{n,n})$ and $n$ by $E(D_{n,n}) = \frac{a}{\sqrt{n}}$, and then estimated how large $n$ should be so that $E(D_{n,n}) \leq 0.01$. The function $\frac{a}{\sqrt{n}}$ was chosen because of the relationship between $E(D_{n,n})$ and $n$ for the i.i.d. case and because it produced a strong fit in all instances (the coefficients of determination, $R^2$, for a least squares fit ranged from 0.9628 to 0.9953). We also fitted power functions that related $Q_{0.05}(D_{n,n})$ and $n$, as well as $Q_{0.95}(D_{n,n})$ and $n$, by $Q_{0.05}(D_{n,n}) = \frac{b}{\sqrt{n}}$ and $Q_{0.95}(D_{n,n}) = \frac{c}{\sqrt{n}}$ (the values for $R^2$ ranged from 0.8901 to 0.9952). Figure \ref{fig:fitn} shows an example of such curve fitting. Table  \ref{table:ctable} shows the estimated values of $n$ that give $E(D_{n,n}) \approx 0.01$ for all seven ranked Democratic vote shares and number of Democratic seats using data from three different races, for chains with weight factors of 1 and 20. 

\begin{table}[h!]
	\centering
	\begin{tabular}{lr:rr:rr:r}
		\hline
		& \multicolumn{2}{c}{Governor} & \multicolumn{2}{c}{Treasurer} & \multicolumn{2}{c}{Secretary of State}  \\ 
		\hline			
		Rank 1 vote share & 
		811106 & 3098768 & 
		806650 & 3116814 & 
		825556 & 3258335 \\  
		Rank 2 vote share  & 
		390844 & 566189 & 
		431378 & 658059 & 
		299628 & 567881 \\ 
		Rank 3 vote share  & 
		567756 & 1462898 & 
		623926 & 2129480 & 
		612522 & 1984180 \\
		Rank 4 vote share  & 
		430815 & 1066183 & 
		454447 & 831971 & 
		442351 & 873455 \\
		Rank 5 vote share  & 
		243815 & 1215242 & 
		287414 & 1272266 & 
		265032 & 1279784 \\
		Rank 6 vote share  & 
		234913 & 474645 & 
		190183 & 428576 & 
		202264 & 474049 \\
		Rank 7 vote share & 
		203264 & 982203 & 
		213385 & 920153 & 
		207917 & 914457 \\
		Number of seats & 
		259908 & 611425 & 
		245961 & 263681 & 
		254780 & 289120 \\
		\hline	
	\end{tabular}
	\caption{\footnotesize Estimated $n$ that gives $E(D_{n,n}) \approx 0.01$ for chains with weight 1 (left) and 20 (right). Ranks 1 through 7 refer to the ranked district vote shares (1 = least Democratic, 7 = most Democratic).}
	\label{table:ctable}
\end{table}


We can observe from Table \ref{table:ctable} that the main differences in convergence rate between chains with weight 1 and 20 are for the district vote shares; more specifically, chains with a weight factor of 20 tended to have slower convergence on  vote shares than chains with a weight factor of 1.  For ensembles with weight 1, the largest $n$ was $825,556$ for the lowest share of Democratic votes, while for ensembles with weight 20, the largest $n$ was $3,258,335$ for the lowest share of Democratic votes. 

\begin{figure}[h!]
	\begin{center}
		\includegraphics[width=4.5in]{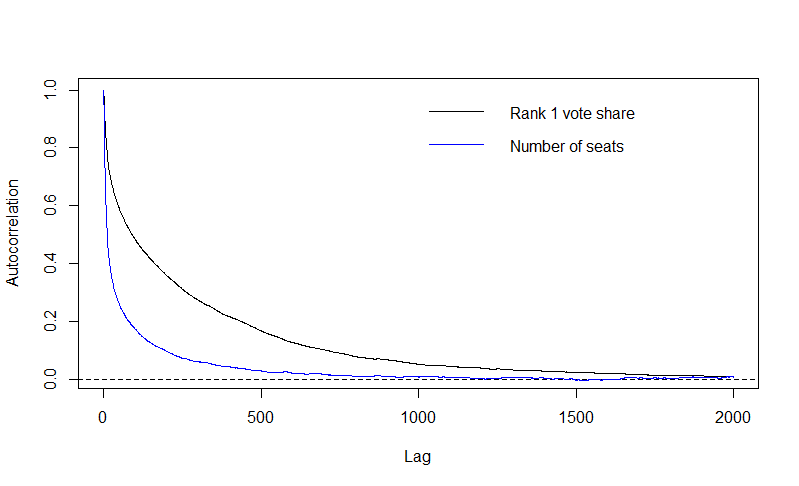}
	\end{center}
	\caption{\footnotesize Autocorrelations for the smallest Democratic vote share and number of Democratic seats in a chain with weight 20 and initial plan E, using data from the Secretary of State's race. In this example, 1962 is the smallest $n$ such that the lag-$n$ autocorrelation for the least Democratic district is less than or equal 0.01.}
	\label{fig:ac}
\end{figure}

The autocorrelations for vote share and number of seats for each of the 10 chains with weight 20 confirmed that the slowest converging measure is the Rank 1 Democratic vote share. More specifically, 1962  was the largest of the smallest values of $n$  (among 10 chains starting from different seeds) such that the lag-$n$ autocorrelation of the least Democratic district was less than or equal to 0.01. Table \ref{table:actable20} shows the range of such lags for all measures, and Figure \ref{fig:ac} shows an example of the rate of decay for the autocorrelations. 

\begin{table}[h!]
	\centering
	\begin{tabular}{ l c c c }
		\hline
		& Governor & Treasurer & Sec. of State \\ 
		\hline
		Rank 1 vote share & 1096 -- 1925 & 1096 -- 1919 & 1107	-- 1962\\  
		Rank 2 vote share & 428	-- 618 & 460 -- 625 & 585	-- 880\\
		Rank 3 vote share & 962 -- 1173 & 1068 -- 1593 & 1073	-- 1495\\
		Rank 4 vote share & 722	-- 1193 & 643 -- 1128 & 684	-- 1145\\
		Rank 5 vote share & 771	-- 1202 & 831 -- 1217 & 832	-- 1215\\
		Rank 6 vote share & 265	-- 430 & 216 -- 294 & 249	-- 318\\
		Rank 7 vote share & 915	-- 1360 & 913 -- 1358 & 913	-- 1358\\
		Number of seats 
		& 822 -- 1136 
		& 583 -- 864
		& 645 -- 919 \\
		\hline					
	\end{tabular}
	\caption{\footnotesize Minimum (left) and maximum (right) values of $n$ (among 10 weight 20 chains starting on different seeds) such that lag-$n$ autocorrelations are less than or equal to 0.01. Ranks 1 through 7 refer to the ranked district vote shares (1 = least Democratic, 7 = most Democratic).}
	\label{table:actable20}
\end{table}

Tables \ref{table:ctable} and \ref{table:actable20} show that, for weight 20 chains, any value of $n$ greater than 3,258,335 will satisfy our sample size criteria. Given that, for large $n$, our data suggests that $E(D_{n,n})$ decreases very slowly as $n$ increases, and given that  only four values on Table \ref{table:ctable} were greater than 2 million (three corresponding to the least Democratic district), we deemed it appropriate to use ensembles of size $n=2,000,000$ for the data analyses and summaries in section \ref{sec:data}. 
In this case, $E(D_{n,n}) \leq 0.0128$ and $n$ is 1019 times as large as the number of lags it takes for the slowest decaying autocorrelation to be 0.01 or less. 
Table \ref{table:dtable20} has the estimates for $E(D_{n,n}),$ $Q_{0.05}(D_{n,n})$ and $Q_{0.95}(D_{n,n})$ for $n=2$ million and a weight factor of 20. These estimates give an idea of the deviations in distribution that might be incurred if one were to run the analyses with ensembles generated by different seeds.

\begin{table}[h!]
	\centering
	\begin{tabular}{ l c c c }
		\hline
		& Governor & Treasurer & Secretary of State \\ 
		\hline
		Rank 1 vote share & 0.0124 [0.0048, 0.0238] & 0.0125 [0.0049,	0.0241] & 0.0128 [0.0048, 0.0247]\\  
		Rank 2 vote share & 0.0053 [0.0025, 0.0093] & 0.0057 [0.0028,	0.0092] & 0.0053 [0.0026, 0.0090]\\
		Rank 3 vote share & 0.0086 [0.0032, 0.0163] & 0.0103 [0.0035,	0.0203] & 0.0100 [0.0035, 0.0196]\\
		Rank 4 vote share & 0.0073 [0.0036, 0.0120] & 0.0064 [0.0032,	0.0104] & 0.0066 [0.0032, 0.0107]\\
		Rank 5 vote share & 0.0078 [0.0033, 0.0144] & 0.0080 [0.0035,	0.0149] & 0.0080 [0.0035, 0.0150]\\
		Rank 6 vote share & 0.0049 [0.0026, 0.0078] & 0.0046 [0.0027,	0.0071] & 0.0049 [0.0026, 0.0076]\\
		Rank 7 vote share & 0.0070 [0.0030, 0.0125] & 0.0068 [0.0029,	0.0120] & 0.0068 [0.0029, 0.0120]\\
		Number of seats 
		& 0.0045 [0.0011, 0.0087] 
		& 0.0030 [0.0008, 0.0061] 
		& 0.0031 [0.0008, 0.0062] \\
		\hline
	\end{tabular}
	\caption{\footnotesize Estimates for $E(D_{n,n})$ and 90\% prediction intervals for $D_{n,n}$, constructed with sample quantiles $[Q_{0.05}(D_{n,n}), Q_{0.95}(D_{n,n})]$, for district vote shares and number of Democratic seats, in chains of size $n=2,000,000$ and a weight factor of 20. Ranks 1 through 7 refer to the ranked district vote shares (1 = least Democratic, 7 = most Democratic).} 
	\label{table:dtable20}
\end{table}


\section{Data and Analysis}\label{sec:data}

In this section, we present the data obtained from our $2,000,000$ step chain with weight factor 20.  For each districting plan in the resulting ensemble, precinct-level voting data was used to evaluate outcomes for that plan, and the results were aggregated to show the range of outcomes across the ensemble.

\subsection{Colorado Redistricting Baseline and Comparison to Enacted Plan}\label{subsec:baseline}
Our first goal was to establish a baseline against which to compare the enacted plan.  Here we will present results from each of the three races we studied (Governor, Treasurer, and Secretary of State).

\begin{figure}[h]
\begin{center}
\includegraphics[width=2in]{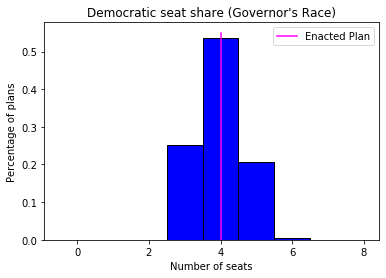}
\includegraphics[width=2in]{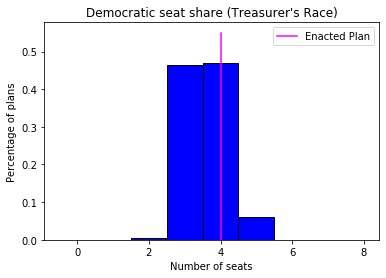}
\includegraphics[width=2in]{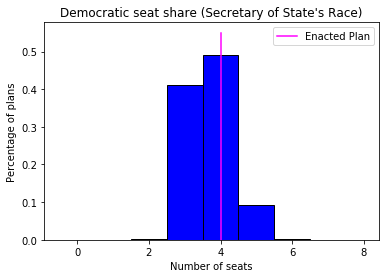}
\end{center}
\caption{\footnotesize Democratic seat share}
\label{Dem-seatshare-W20-fig}
\end{figure}

{\bf Democratic seat share:} Figure \ref{Dem-seatshare-W20-fig} shows histograms for the number of Democratic seats won for each plan in the ensemble, based on voting data from each race.  The Democratic seat share for the enacted plan for each race is also shown for comparison.  For all three races, the enacted plan produced 4 Democratic seats; moreover, this outcome was the most common result across all plans in the ensemble for all three races.

\begin{figure}[h]
\begin{center}
\includegraphics[width=2in]{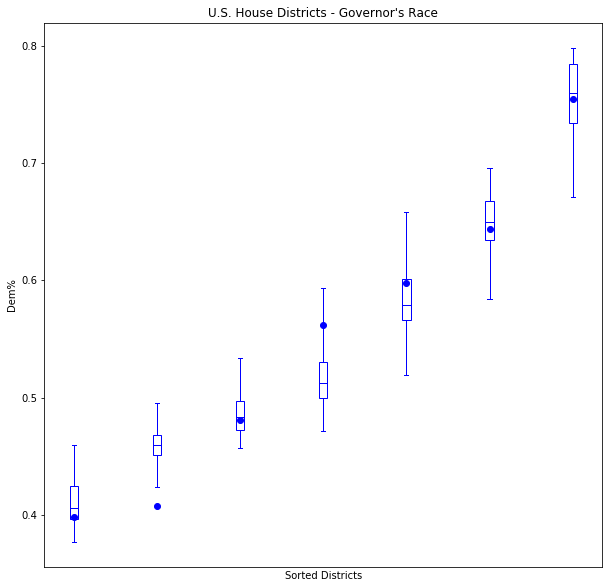}
\includegraphics[width=2in]{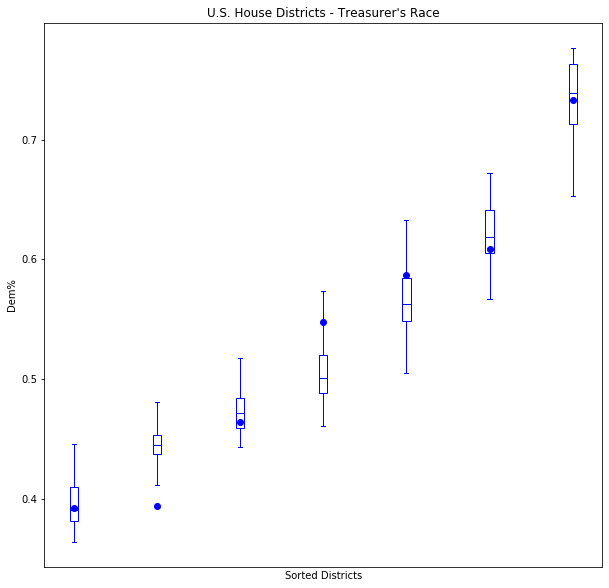}
\includegraphics[width=2in]{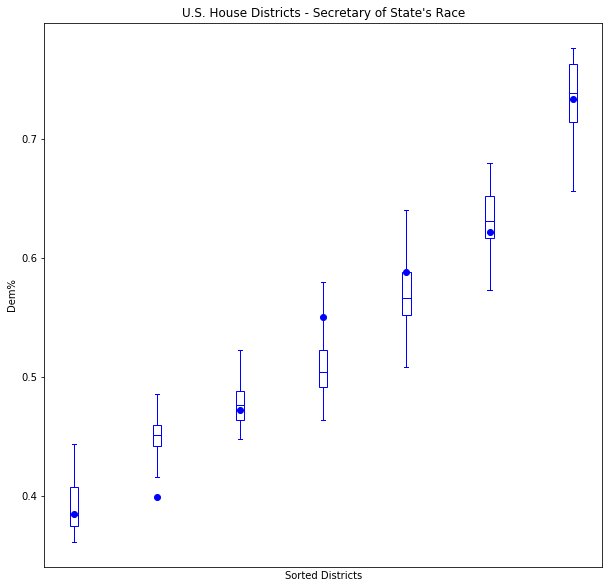}
\end{center}
\caption{\footnotesize Districts sorted by Democratic vote share}
\label{Dem-boxplots-W20-fig}
\end{figure}

{\bf Democratic vote share by district:}
Figure \ref{Dem-boxplots-W20-fig} shows a more detailed picture of the partisan results of these elections. For each plan in the ensemble, districts are sorted by Democratic vote percentage, from lowest to highest.  The box plots show the range of Democratic vote shares in the sorted districts (so, e.g., the second box from the left shows the range of Democratic vote shares in the second most Republican district in each plan), with the boxes showing the middle 50\% and the whiskers extending from the $1$st percentile through the $99$th.  Here we see in more detail how the enacted plan compares to the ensemble: For each of these elections, the second district has significantly fewer Democratic votes than might be expected---fewer than over $99\%$ of the ensemble. 
Meanwhile, the fourth district has a higher than average Democratic percentage---somewhere around the middle of the top quartile.

\begin{figure}[h]
\begin{center}
\includegraphics[width=3in]{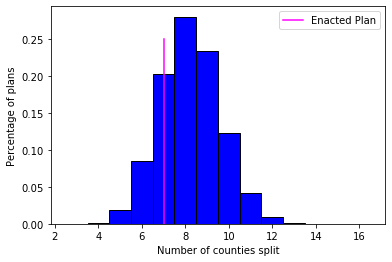}
\includegraphics[width=3in]{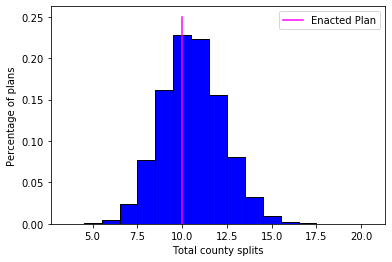}
\end{center}
\caption{\footnotesize Total county splits}
\label{county-splits-fig}
\end{figure}

{\bf County splits:}
The first plot in Figure \ref{county-splits-fig} shows a histogram for the numbers of counties that are split into multiple districts in each plan.  The second plot shows a histogram for a variation on this measure: For every county that is split into more than 1 district, this measure counts 1 less than the number of districts that contain a portion of the county.  (So, e.g., if portions of a county are contained within 3 different districts, then it counts as 1 county split in the first plot and as 2 total splits in the second plot.)  For both measures, the enacted plan has one fewer split than the mode of the ensemble, but is still well within the expected range. 

\begin{figure}[h]
\begin{center}
\includegraphics[width=3in]{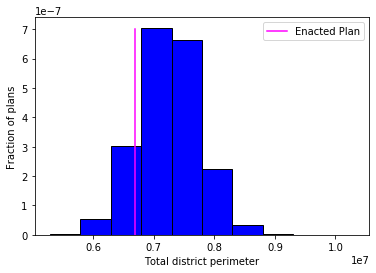}
\end{center}
\caption{\footnotesize Total district perimeter}
\label{weight20-perimeter-fig}
\end{figure}

{\bf Compactness:} While Amendments Y and Z do not give a specific definition for district compactness, Colorado has historically considered the total perimeter of all districts in a districting plan as a measure of compactness and has sought to minimize this quantity.  Figure \ref{weight20-perimeter-fig} shows a histogram for total district perimeter for each plan in the ensemble, with the enacted plan included for comparison.  As we can see from this figure, the enacted plan is well within the expected range, and slightly more compact by this measure than the mean for the ensemble.

\subsection{Competitiveness}\label{subsec:competitiveness}
Competitiveness of districts is widely regarded as desirable, but the notion of ``competitiveness" is not so easy to quantify.  
A variety of competitiveness metrics were explored in \cite{DDS20}, and the analysis given there shows that there may be situation-specific difficulties with creating good competitiveness metrics, and that optimizing competitiveness metrics may produce unintended consequences.  
Recall that for purposes of our analysis, we have defined a district to be competitive with respect to a particular election if the Democratic vote share for that district (ignoring votes for minor parties) is between 45\% and 55\%.  These margins are certainly somewhat arbitrary, and in light of the results of \cite{DDS20}, we advise caution in the interpretation of these results.  
In particular, as we shall see below, a significant change in the overall partisan outcome (for instance, a ``wave year" for one party or the other) can completely change the characterization of a district as competitive or not by this measure.

\begin{figure}[h]
\begin{center}
\includegraphics[width=2in]{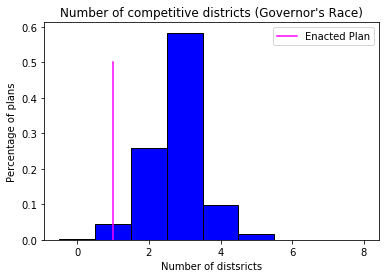}
\includegraphics[width=2in]{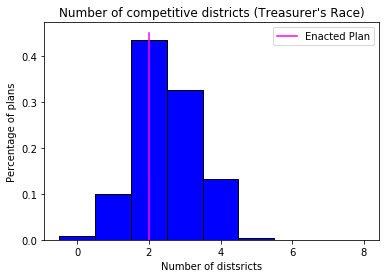}
\includegraphics[width=2in]{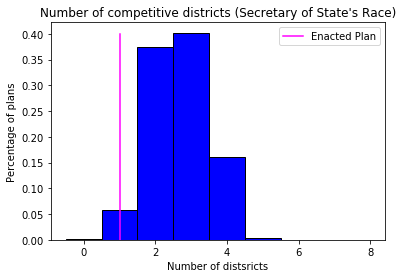}
\end{center}
\caption{\footnotesize Competitive districts}
\label{competitiveness-actual-fig}
\end{figure}

Figure \ref{competitiveness-actual-fig} shows histograms for the number of competitive districts for each plan in the ensemble, based on voting data from each race.  The number of competitive districts for the enacted plan for each race is also shown for comparison.  Note that for the enacted plan, the Governor and Secretary of State elections each produced 1 competitive district, which appears to be something of an outlier compared to the ensemble.  However, the Treasurer election produced 2 competitive districts, which was the most common outcome in the ensemble for that race.  

Referring back to Figure \ref{Dem-boxplots-W20-fig}, we see that the 3rd-least Democratic district in the enacted plan falls within the competitive range for all three elections (48.01\%, 46.43\%, and 47.18\%).  The middle district is just barely within the competitive range for the Treasurer election (54.77\%) and just outside the competitive range for the Governor and Secretary of State elections (56.17\% and 55.06\%, respectively).  Meanwhile, the overall statewide Democratic percentages for these races were 55.52\% (Governor), 53.76\% (Treasurer), and 54.11\% (Secretary of State).  

From these results, we can already see that the number of competitive districts is an extremely delicate measure.  The somewhat arbitrarily selected range of 45\% - 55\% creates a scenario where, for instance, a difference of 0.06\% in one district in the Secretary of State election is the difference between 1 competitive district (an apparent outlier) and 2 competitive districts (squarely the in the main portion of the histogram).  Moreover, this measure is extremely sensitive to the overall partisan outcome of the election; even among these elections that were held on the same day, we see that the Treasurer election had the overall partisan split that was closest to 50-50 and was also the only election to produce 2 competitive districts.  In a different election with a different partisan lean, we might expect completely different results.

In order to explore these ideas more fully, we used the notion of ``uniform partisan swing" to normalize each of these elections to create a fictional election 
with 50-50 partisan split.  
This is a standard construction in political science that is used for generating the ``seats-votes curves" that give rise to traditional measures of partisan symmetry; see, e.g., \cite{DDDGMSS21}.
So, for instance, for the Governor election that originally had a 55.52\% Democratic percentage, we subtracted 5.52\% from the Democratic vote share in each district for each plan in the ensemble.  This has the effect of shifting each of the box plots in the first graph in Figure \ref{Dem-boxplots-W20-fig} down by 5.52\%.

\begin{figure}[h]
\begin{center}
\includegraphics[width=2in]{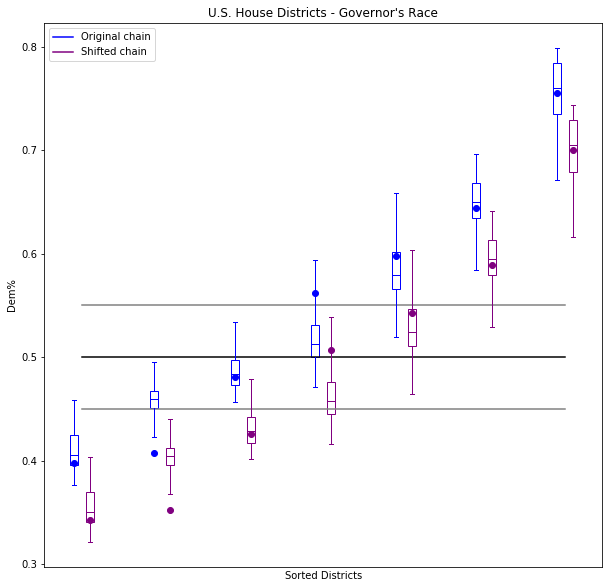}
\includegraphics[width=2in]{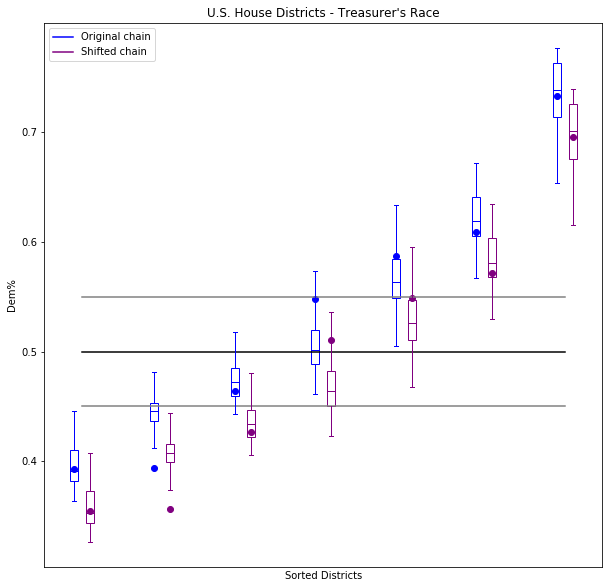}
\includegraphics[width=2in]{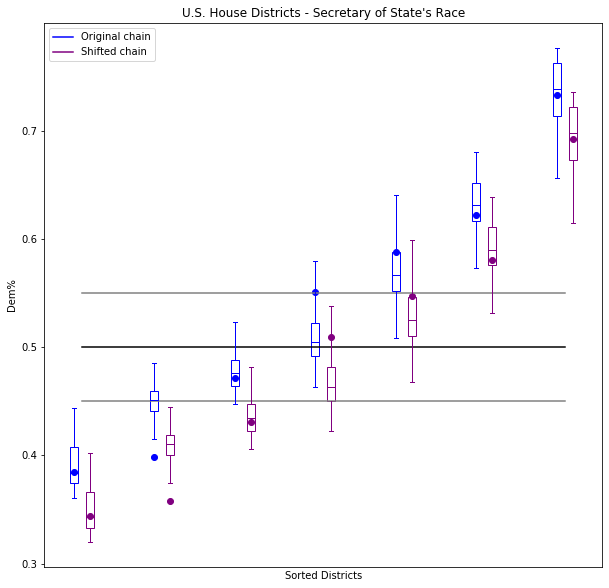}
\end{center}
\caption{\footnotesize Sorted districts, original and shifted}
\label{Dem-boxplots-W20-shifted-fig}
\end{figure}

In Figure \ref{Dem-boxplots-W20-shifted-fig}, we show the box plots from Figure \ref{Dem-boxplots-W20-fig} side by side with the box plots for the hypothetical elections with 50-50 partisan split.  For reference, these graphs also show horizontal lines at 45\%, 50\%, and 55\% so that we can easily see which districts lie in the competitive range.
The picture tells an interesting story: for the original elections, the 2nd, 3rd, and 4th districts appear to be the most competitive, while for the hypothetical 50-50 elections, the 4th and 5th districts are the most competitive.

\begin{figure}[h]
\begin{center}
\includegraphics[width=2in]{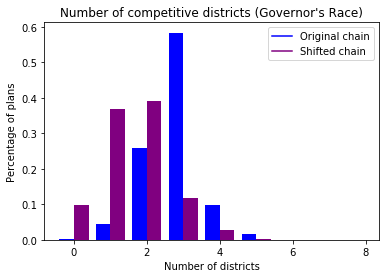}
\includegraphics[width=2in]{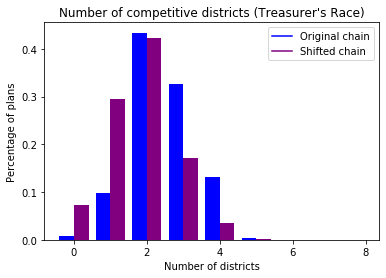}
\includegraphics[width=2in]{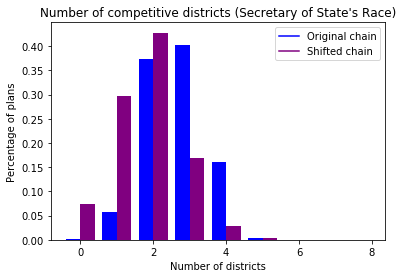}
\end{center}
\caption{Competitive districts, original and shifted}
\label{competitiveness-swung-fig}
\end{figure}

Similarly, in Figure \ref{competitiveness-swung-fig} we plot the histograms from Figure \ref{competitiveness-actual-fig} for the numbers of competitive districts for each race side by side with the same histograms for the 50-50 elections.  As we gathered from the plots in Figure \ref{Dem-boxplots-W20-shifted-fig}, the hypothetical 50-50 elections actually produce {\em fewer} competitive districts than the original elections.  While these results may seem counterintuitive---shouldn't an election with a partisan split closer to 50-50 have more competitive districts than a 55-45 election?---they serve to highlight that competitiveness is a more complicated issue than it might appear at first glance.

\subsection{County Splits}\label{subsec:splits}
One important question for redistricting is how various priorities interact with each other; for instance, how does a focus on preserving communities of interest impact partisan seat share and/or the ability to draw competitive districts?
In order to explore this question, in this section we examine how the numbers of county splits (both number of counties split and total county splits) are related to partisan outcomes and to the number of competitive districts.

\begin{figure}[h]
\begin{center}
\includegraphics[width=3in]{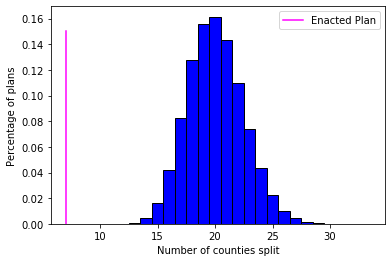}
\includegraphics[width=3in]{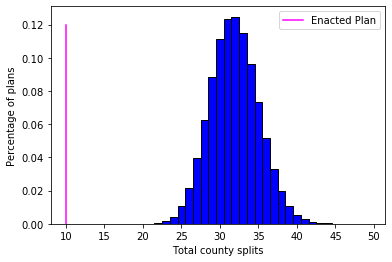}
\end{center}
\caption{\footnotesize Total county splits}
\label{unweighted-county-splits-fig}
\end{figure}

Figure \ref{county-splits-fig} above shows the distributions of county splits for our weight 20 chain.  We also ran a 2,000,000 step Markov chain using the standard, unweighted (i.e., weight 1) ReCom method, and the distributions of county splits for this chain is shown in Figure \ref{unweighted-county-splits-fig}.  Since the primary difference between these two chains is the weighting algorithm used to reduce the number of county splits in the weight 20 chain, it seems reasonable to hypothesize that any significant differences between other metrics for these two chains are indicative of a relationship between these metrics and county splits.

\begin{figure}[h]
\begin{center}
\includegraphics[height=1.1in]{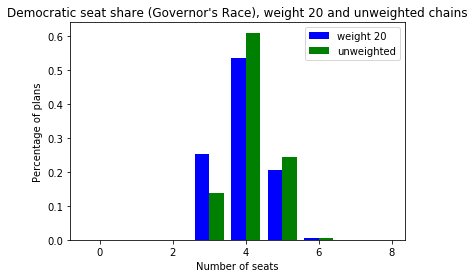}
\includegraphics[height=1.1in]{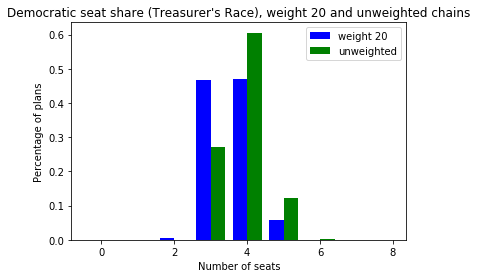}
\includegraphics[height=1.1in]{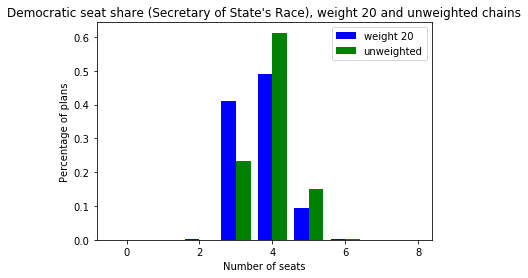}
\end{center}
\caption{\footnotesize Democratic seat share for weight 20 and unweighted chains}
\label{Dem-seatshare-compare-fig}
\end{figure}

In Figure \ref{Dem-seatshare-compare-fig}, we plot the histograms for Democratic seat share for each race from the weight 20 chain side by side with the same histograms for the unweighted chain.  From these histograms, we see that the unweighted chain tends to produce slightly more Democratic seats than the weight 20 chain.  

More quantitatively, the mean Democratic seat shares are shown in Table \ref{Dem-seatshare-compare-table}.  The effect of constraining the numbers of county splits in the weight 20 chain appears to be that the mean Democratic seat share is reduced by (depending on the race) approximately $1/4$ of a seat compared to the unweighted chain. 

\begin{table}[h]
	\centering
	\begin{tabular}{ l |c |c}
		\hline
		Race & Mean Dem seat share (weight 20) & Mean Dem seat share (unweighted)  \\ 
		\hline
		Governor & 
		3.96 (0.006) & 
		4.12 (0.003) \\  
		Treasurer & 
		3.58 (0.004) & 
		3.85 (0.003) \\
		Secretary of State & 
		3.68 (0.004) & 
		3.92 (0.003)
	\end{tabular}
	\caption{\footnotesize Mean Democratic seat shares in weight 20 and unweighted chains. Standard error estimates (in parentheses) were calculated using ten weight 20 chains and ten unweighted chains starting at different plans. }
	\label{Dem-seatshare-compare-table}
\end{table}

\begin{figure}[h]
\begin{center}
\includegraphics[width=3in]{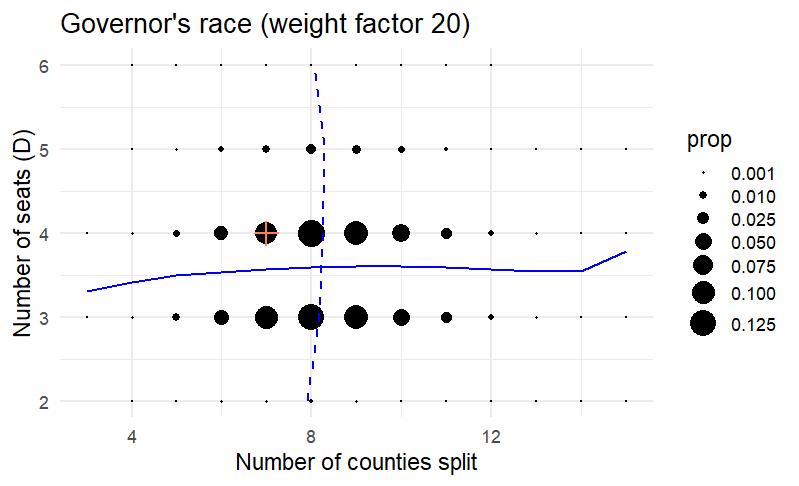}
\includegraphics[width=3in]{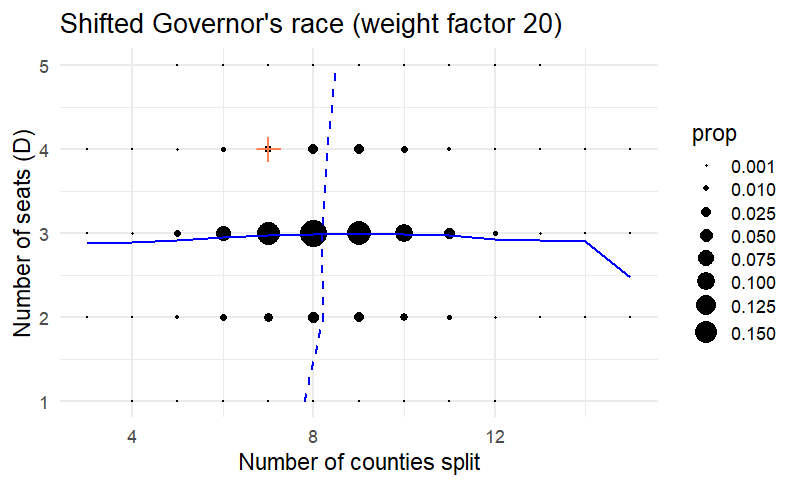}
\end{center}
\caption{\footnotesize Democratic seats vs. county splits for Governor's race in weight 20 chain}
\label{Dem-seats-vs-splits-within-chain-fig}
\end{figure}

We also looked at the relationship between county splits/total splits and Democratic seat share within each chain, where the range of splits is much smaller than the difference between the weighted and unweighted chains.  Perhaps not surprisingly, considering the small magnitude of the difference in Democratic seats between the two chains, we found very little variation in the number of Democratic seats as the number of county splits/total splits varies.  For example, Figure \ref{Dem-seats-vs-splits-within-chain-fig} shows scatterplots for the number of Democratic seats vs. the number of county splits for the Governor's race in the weight 20 chain, in both the original and shifted versions.  The solid blue line tracks the mean number of Democratic seats as a function of the number of county splits, while the dotted blue line tracks the mean number of county splits as a function of the number of Democratic seats. The location of the enacted plan in each plot is shown as a cross. 

\begin{figure}[h]
\begin{center}
\includegraphics[height=1.1in]{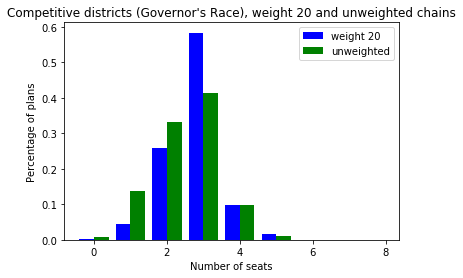}
\includegraphics[height=1.1in]{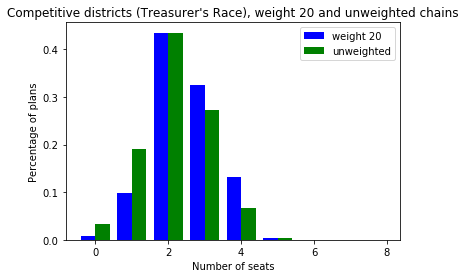}
\includegraphics[height=1.1in]{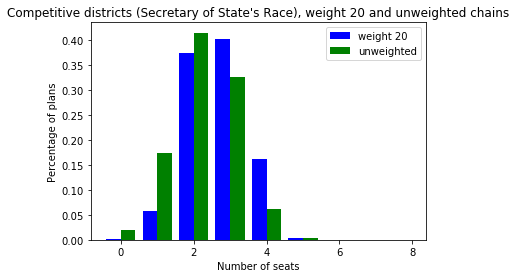}
\end{center}
\caption{\footnotesize Competitive districts for weight 20 and unweighted chains}
\label{competitiveness-actual-compare-fig}
\end{figure}

We can perform a similar comparison for the numbers of competitive districts within these two chains.
In Figure \ref{competitiveness-actual-compare-fig}, we plot the histograms for the number of competitive districts for each race from the weight 20 chain side by side with the same histograms for the unweighted chain.  The means for each race and each chain are shown in Table \ref{comp-dists-compare-table}.  Here we see that the effect of constraining the number of county splits in the weight 20 chain is to increase the number of competitive districts by (depending on the race) approximately $1/3$ of a seat compared to the unweighted chain.

\begin{table}[h]
	\centering
	\begin{tabular}{ l |c |c}
		\hline
		Race & Mean \# competitive (weight 20) & Mean \# competitive (unweighted)  \\ 
		\hline
		Governor & 
		2.78 (0.005) & 
		2.49 (0.003)\\  
		Treasurer & 
		2.48 (0.008) & 
		2.16 (0.004) \\
		Secretary of State & 
		2.67 (0.008) & 
		2.25 (0.003)
	\end{tabular}
 	\caption{\footnotesize Mean number of competitive districts in weight 20 and unweighted chains. Standard error estimates (in parentheses) were calculated using ten weight 20 chains and ten unweighted chains starting at different plans.}
	\label{comp-dists-compare-table}
\end{table}

\begin{figure}[h]
\begin{center}
\includegraphics[height=1in]{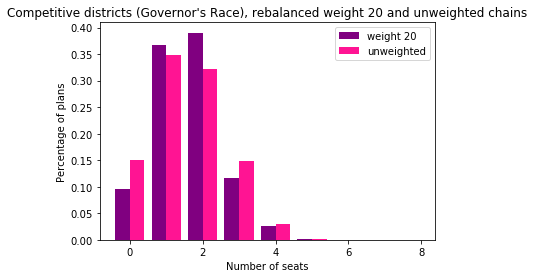}
\includegraphics[height=1in]{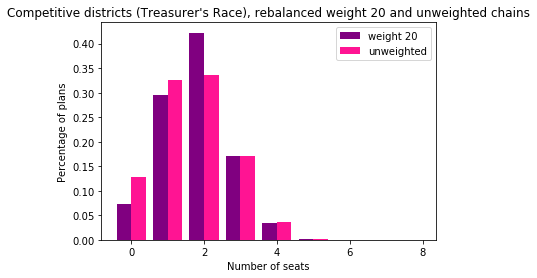}
\includegraphics[height=1in]{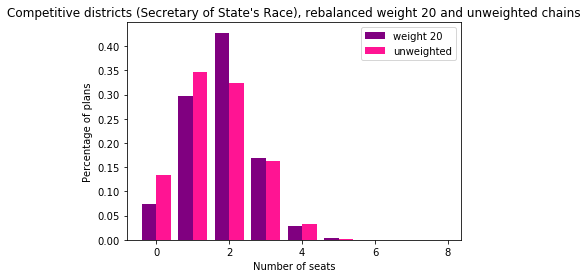}
\end{center}
\caption{\footnotesize Competitive districts for shifted weight 20 and unweighted chains}
\label{competitiveness-swung-compare-fig}
\end{figure}

Since we saw that the number of competitive districts can be extremely sensitive to the overall partisan balance of the election in ways that are not always obvious, we also compare the number of competitive districts for these two chains after performing a uniform swing on each election in each chain to create fictional elections with 50-50 partisan split.  Histograms for the shifted elections for both the weight 20 and unweighted chains are shown in Figure \ref{competitiveness-swung-compare-fig}, with the means for each race and each chain shown in Table \ref{comp-dists-swung-compare-table}. Here we see that the weight 20 chain still has slightly more competitive districts in each race than the unweighted chain, but the effect is smaller after performing the uniform partisan swing.

\begin{table}[h]
	\centering
	\begin{tabular}{ l |c |c}
		\hline
		Race & Mean \# competitive (weight 20) & Mean \# competitive (unweighted)  \\ 
		\hline
		Governor & 
		1.62 (0.008) & 
		1.56 (0.004) \\  
		Treasurer & 
		1.81 (0.008) & 
		1.67 (0.004) \\
		Secretary of State & 
		1.79 (0.008) & 
		1.62 (0.004)
	\end{tabular}
	\caption{\footnotesize Mean number of competitive districts in shifted weight 20 and unweighted chains. Standard error estimates (in parentheses) were calculated using ten weight 20 chains and ten unweighted chains starting at different plans.}
	\label{comp-dists-swung-compare-table}
\end{table}

\begin{figure}[h]
\begin{center}
\includegraphics[width=3in]{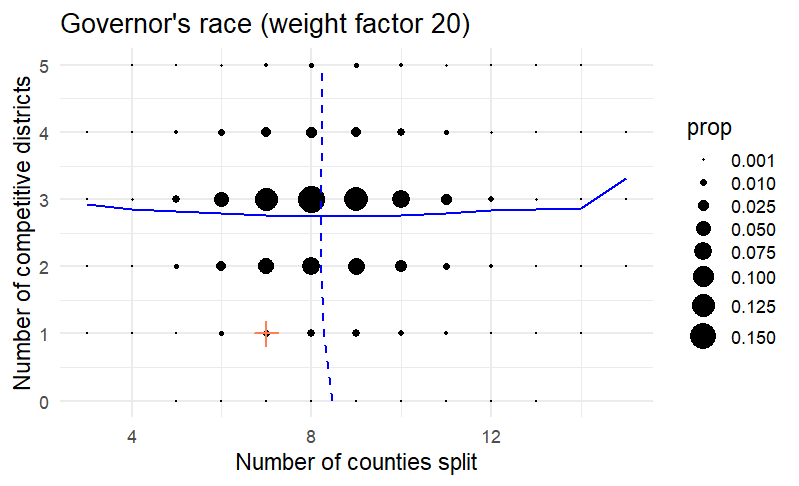}
\includegraphics[width=3in]{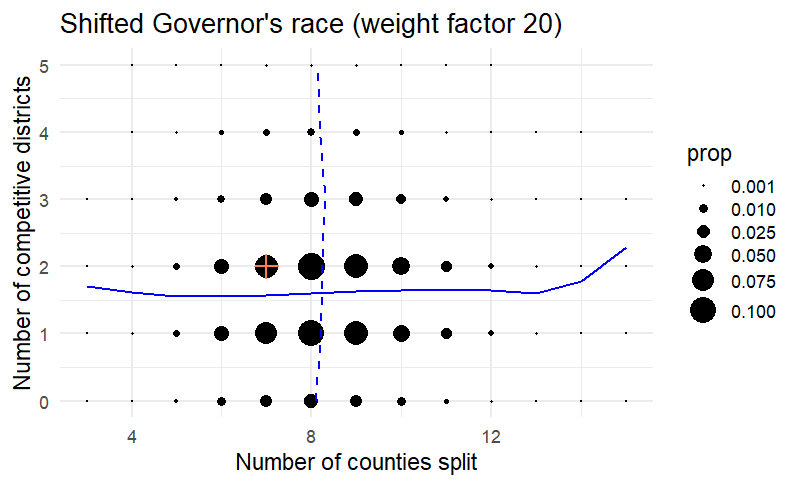}
\end{center}
\caption{\footnotesize Competitive districts vs. county splits for Governor's race in weight 20 chain.}
\label{comp-dists-within-chain-fig}
\end{figure}

We also looked at the relationship between county splits/total splits and competitive districts within each chain, and once again we found very little variation in the number of competitive districts as the number of county splits/total splits varies.
For instance, Figure \ref{comp-dists-within-chain-fig} shows scatterplots for competitive districts vs. the number of county splits for the Governor's race in the weight 20 chain, in both the original and shifted versions.  
The solid blue line tracks the mean number of competitive districts as a function of the number of county splits, while the dotted blue line tracks the mean number of county splits as a function of the number of competitive districts. The location of the enacted plan in each plot is shown as a cross. 

\begin{figure}[h]
\begin{center}
\includegraphics[width=3in]{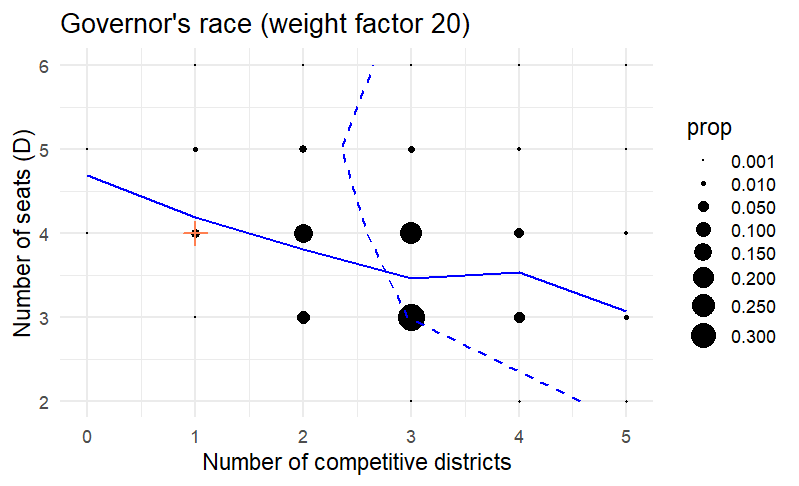}
\includegraphics[width=3in]{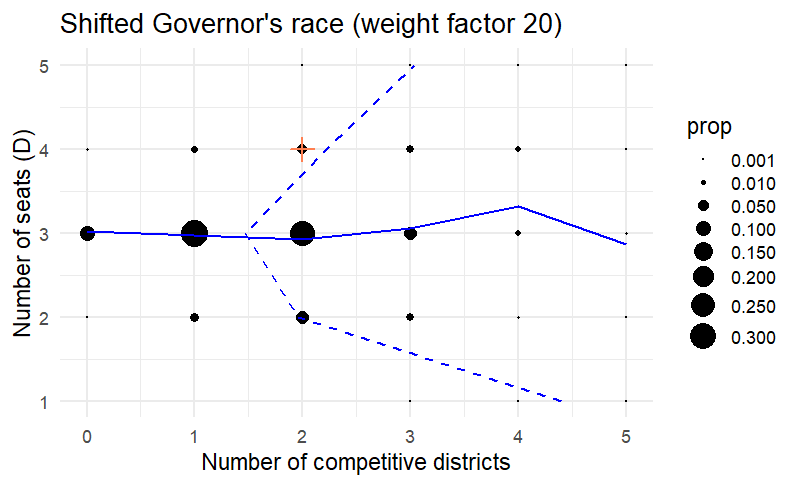}
\end{center}
\caption{\footnotesize Democratic seats vs. number of competitive districts for Governor's race in weight 20 chain.}
\label{Dem-seats-vs-comp-dists-within-chain-fig}
\end{figure}

Finally, we looked at the relationship between the Democratic seat share and the number of competitive districts within each chain.  As an example, Figure \ref{Dem-seats-vs-comp-dists-within-chain-fig} shows scatterplots for the number of Democratic seats vs. the number of competitive districts for the Governor's race in the weight 20 chain, in both the original and shifted versions.  The solid blue line tracks the mean number of Democratic seats as a function of the number of competitive districts, while the dotted blue line tracks the mean number of competitive districts as a function of the number of Democratic seats. The location of the enacted plan in each plot is shown as a cross.

These plots show evidence of an interesting relationship: as we vary the number of competitive districts--a choice that is somewhat under the control of mapmakers---the mean number of Democratic seats changes very little.  But if we consider these plots in the other direction, we see that plots with an extreme number of Democratic seats---either significantly more or less than the mean---tend to have more competitive districts.  We will discuss this observation in more detail in the following section.

\section{Conclusions}\label{sec:conclusions}
The data and discussion in Section \ref{subsec:baseline} paint a picture of a likely baseline of outcomes for Congressional districting plans in Colorado in the 2018 elections. Here, we address the other focus questions from Section \ref{subsec:synthesis}.

\subsection{Do we find evidence in the 2018 elections of effective partisan manipulation in 2011 redistricting?}

The short answer here is that we do not find evidence of effective partisan manipulation in the 2011/2012 adopted maps.  The number of seats won by Democrats under the enacted plan is in no way an outlier--it is in fact the most common outcome for every race we examined under realistic county splitting constraints, as is depicted in Figure \ref{Dem-seatshare-W20-fig}, and this outcome is even more common if more county splits are allowed (see Figure \ref{Dem-seatshare-compare-fig}).  

There is, initially at least, some cause for potential concern about packing Republicans into the second-least Democratic district based on Figure \ref{Dem-boxplots-W20-fig}.  Only 
0.06\% 
of the Monte Carlo sample was below the enacted second-ranked Democratic district in the Governor's race. This number was
0.03\% 
for the Treasurer's race and 
0.09\%
for the Secretary of State's race.
Taking into account all districts, the estimated probability that at least one district (out of the seven) is at the bottom or top 0.06\% of the possible Democratic vote share for the Governor's race is 
0.008.
For the Treasurer's race, the estimated  probability that at least one district is in the bottom or top 0.03\% is 
0.004.
Doing the same for Secretary of State, this probability was 
0.012.
So for all races, the vote shares of the enacted plan for the second-ranked district seems to be an unusual occurrence.

In investigating the enacted district in question, we discovered that, surprisingly, the second-ranked district depends on the race. In Table \ref{DistrictVoteShares}, we see that CO 5 is the second-ranked district in the Governor and Treasurer's races, while CO 4 is the second-ranked district in the Secretary of State's race.  In all cases, the vote shares in the first-ranked and second-ranked districts are extremely close.  Thus this is not a simple case of packing Republicans into a single district.

\begin{table}
\begin{footnotesize}
\begin{center}
\begin{tabular}{l|c|c|c}
District & Governor & Treasurer & Sec. of State\\
\hline
CO 1 & 0.755025 & 0.732848 & 0.733744\\
CO 2 & 0.644199 & 0.608735 & 0.621864\\
CO 3 & 0.480772 & 0.464260 & 0.471826\\
CO 4 & 0.397891 & 0.394212 & 0.398713\\
CO 5 & 0.407602 & 0.392769 & 0.384713\\
CO 6 & 0.561725 & 0.547696 & 0.550642\\
CO 7 & 0.597754 & 0.586725 & 0.588472
\end{tabular}
\end{center}
\end{footnotesize}
\caption{Democratic vote shares for enacted districts in each of the three 2018 focus races.}
\label{DistrictVoteShares}
\end{table}

District CO 4 can be broadly characterized as the Eastern Plains, while CO 5 is El Paso county and the more rural counties just to the west. In the 2011/2012 redistricting, the boundaries of CO 5 changed very little, and then only to move a whole county (Lake) for population balance.  District CO 4 lost Larimer county, and gained the eastern rural parts of Douglas, Adams, and Arapaho counties, and all of Otero, Elbert, and Los Animas counties. The maps were drawn by Democrats, after the redistricting process moved to litigation when the state legislature could not agree on a plan. 
 
The map was chosen out of several options advanced by the parties and other interest groups, because it ``most accurately reflected and preserved current communities of interest in 2011" as well as producing the ``maximum amount of competition of any of the realistically proffered maps in at least three districts--the 3rd, the 6th and the 7th" \cite{DPMoreno}.  Larimer County was moved to CO 2 because Fort Collins (site of the main Colorado State University campus)  was deemed to have a shared community of interest with Boulder (home of University of Colorado's largest campus) in higher education, health care, and tech.  The rationale for adding the other areas to CO 5 was to bring population balance and preserve communities of interest in agriculture and oil/gas/fracking concerns. 

One way to look at these results is to conclude that, in aggregate, districts CO 4 and CO 5 are unusually Republican in the enacted plan.  This certainly has the effect of putting more Democrats in other districts.  However, the district CO 2, where Larimer county moved, was previously a reliably Democratic district, so this change had little effect on the partisan lean of CO 2, CO 4, or CO 5.  

The only other somewhat unusual-appearing enacted district is the less extreme over-concentration of Democratic voters in the fourth-ranked district.  While not an outlier, the seat share of the enacted district is outside the middle 50\% of the ensemble.  However, we note that having some district outside of the middle 50\% on its own would not be at all an unusual occurrence.  If the district seat shares were independent draws, there would only be a $\frac{1}{2^7}$ probability of having all the seven districts in this middle range.  In fact, the district seat shares are not independent, so one would expect that with CO 4 and CO 5 having unusually few Democrats, some other collection of districts must have more than the median number of Democrats.  The fourth-ranked district is CO 7, containing the northwest suburbs of Denver.  When the map was adopted CO 7 was considered a competitive district, and the district with this core was held by Democrats both before and after the redistricting process.  

This situation draws attention to the fact that respecting communities of interest and seeking competitive districts may naturally result in grouping and dividing voters in a way that could be called packing and cracking, regardless of partisan motivation. In all cases some voters will be unhappy with districting lines.  We conclude that while the enacted plan does display some features which in some contexts are indicative of packing and cracking, the partisan outcome of the enacted plan is so thoroughly in line with the ensemble that this cannot be taken as evidence of successful partisan gerrymandering.

\subsection{Do we find evidence among our generated plans that the number of county splits in a plan is related to partisan outcomes?}

Within the range of county splits that have historically been deemed acceptable in enacted maps, the number of county splits is fairly independent of partisan outcome.  However, we also compare the ``weighted" ensemble of plans with realistically few county splits to the ``unweighted" ensemble, which tends to have many more splits.  We find a small tendency of about $0.25$ more Republican seats for the weighted compared to the unweighted.

Some political scientists, including Colorado politics and government specialist Robert Loevy, have advocated ignoring county boundaries in drawing lines, arguing that counties are not meaningful communities of interest and that breaking county boundaries could allow more competitive maps to be drawn \cite{Loevy2011}. We conclude that, at this time in Colorado, ignoring county boundaries in map drawing would slightly benefit Democrats on average. However, the range of non-outlier partisan outcomes which occur in the unweighted ensemble is the same as that range for the weighted, so plans identified as extreme partisan gerrymanders in a weighted ensemble would not appear reasonable within an unweighted ensemble, either.

\subsection{Do we find evidence among our generated plans that the number of competitive districts in a plan is related to partisan outcomes?}

If we sort all plans in a given ensemble by the number of competitive seats in the plan, the average number of Democratic seats for each group seems to vary only slightly depending on the number of competitive seats.  For weighted and unweighted ensembles with actual voting data, there is a slight trend to have fewer Democratic seats as the number of competitive seats is increased.  This trend does not exist when using shifted voting data.  When we instead sort all plans in a given ensemble by the number of Democratic seats, we see a clearer trend, 
as seen in Figure \ref{Dem-seats-vs-comp-dists-within-chain-fig}.
First, within the range of Democratic seats that could be deemed common outcomes (i.e. 3, 4, or 5 Democratic seats in both the weighted and unweighted ensembles), there is a small trend of increasing average number of competitive districts as the number of Republican seats increases.  This could simply reflect the fact that in this election, Republican voters were in the minority in Colorado.  Thus among the plans that were better for Republicans, we would need to see most districts which Republicans won having having relatively narrow Republican majorities, meaning that they would count as competitive districts.  This hypothesis is backed up by the fact that there is no discernible trend when we use shifted voting data, leaving neither party in the minority.  This leads us to conjecture that maximizing (or minimizing) the numbers of competitive seats in a plan would tend to generally result in slightly better outcomes for the minority party, especially within the most common range of competitive seats, but that this is not a strong trend.

We next observe that plans with uncommonly many or few Democratic seats tended to have more competitive seats.  Upon reflection, this makes sense because, again based on the total number of voters of each party within the state, more extreme outcomes will require more close races.  If a party is to win many races in a state with relatively competitive overall vote shares, that party cannot win by a large margin in many districts. If one were to engineer non-competitive districts, this would also constrain the range of possible outcomes. We see that for (the very few) plans with 0 competitive districts in the weighted ensemble, we observe outcomes of 4, 5, or 6 Democratic seats, reflecting the fact that it is very difficult to engineer more than 3 safe Republican seats or fewer than 1 safe Democratic seat.  For plans with 1, 2, or 3 competitive districts we see outcomes ranging from 3-6 seats, and for 4 or 5 competitive districts we see outcomes from 2-6 seats.  This tendency of spread in partisan outcomes for plans with more competitive seats and the higher average number of competitive seats for more extreme partisan outcomes hold for both weighted and unweighted ensembles and for shifted voting data in both cases.  Thus while we do not conclude that there is aggregate benefit to either party in creating more competitive districts, one might look among plans with more competitive districts if one desired a more extreme partisan outcome in either direction.  This could allow unscrupulous map-makers to justify extreme partisan outcomes by claiming that their map was engineered to be extremely competitive.

\subsection{Do we find evidence among our generated plans that the number of competitive districts in a plan is related to the number of county splits in that plan?}

In all ensembles, we see that within the range of that single ensemble, there is no apparent relationship between the number of competitive districts the number of county splits, or the number of total counties split.  However, when we compare the weighted and unweighted ensembles, we see an average of around $0.3$ more competitive districts when we restrict county splits compared to allowing free county splitting.  This is perhaps surprising, because it would seem that a reasonable way of maximizing competitiveness would be to carve up ``strongholds" of both parties--for example, splitting heavily Democratic Denver county into pieces containing swaths of the eastern plains or western slope, and breaking Republican-leaning El Paso county to combine with more Democratic Pueblo and the southern suburbs of Denver.  Indeed, there is some evidence that for a human map-maker working with very similar Colorado data, preserving county boundaries can interfere with maximizing the number of competitive districts, and the interference is greater than it would be if the map-maker attempted to create compact districts instead of preserving county boundaries \cite{Liller}.  We conclude that it is in fact not necessary to sacrifice competitiveness to preserve county boundaries, or to break a large number of county boundaries to get a reasonably large number of competitive districts.

\section{Further Questions}\label{sec:questions}
We conclude with a few questions and directions for further study that emerge from this work.
\begin{enumerate}
\item What are the electoral baselines for the Colorado State Legislature?  What are the answers to the focus questions in that context?  Though an independent commission was responsible for the 2011/2012 Colorado legislative redistricting, the process was also extremely contentious and the outcome was again controversial, so the question of effective manipulation is again relevant. The fact that there are many more districts for each chamber of the legislature would allow for a much finer analysis of the relationships between fairness criteria, as well. 
\item Why do we see more competitive plans when county splits are reduced?  Is this a universal principle, or specific to Colorado?  We hypothesize that some of this effect is due to the fact that with larger ``building blocks" (counties instead of precincts), the blocks tend to have less extreme partisan vote shares, and thus districts built of these blocks cannot reach the extremes that might be possible using smaller blocks.
\item What factors contribute to the determination of a reasonable ensemble size, and how can this size be estimated in general?  We expect that this number will be affected by factors such as the number of districts in the plan, the number of underlying units (e.g., precincts), and the variance of vote data across units; moreover, these factors may interact in ways that have not yet been explored.
\item How does varying the weight parameter change the distribution of sampled plans?  We have done only a preliminary analysis of this parameter's effects.
\item What is a meaningful and useful definition of competitiveness in this context?  
\end{enumerate}

\section{Acknowledgments}
We would like to thank University of Nebraska graduate student Austin Eide for invaluable assistance in getting started; Todd Bleess of the Colorado State Demography Office for a great starting map; Geographers Dr. Rebecca Theobald and Dwayne Liller; student researchers Edgar Santos Vega, Jose Monge Castro, and Kadin Mangalik; and generous Colorado College GIS experts Matt Cooney and Francis Russell.

\section{Conflict of Interest Statement}
On behalf of all authors, the corresponding author states that there is no conflict of interest.

\nocite{baseline}
\nocite{PA-report}

\bibliographystyle{amsplain}
\bibliography{CO-Ensemble-Analysis-bib}

\end{document}